\documentclass{article}

\usepackage{fullpage,verbatim}
\usepackage{theorem}
\usepackage{amsmath,amsfonts,amssymb,graphicx}

\newtheorem{definition}{Definition}
\newtheorem{proposition}{Proposition}
\newtheorem{corollary}{Corollary}
\newtheorem{theorem}{Theorem}
\newenvironment{proof}{\noindent\textbf{Proof}}{\hfill$\Box$}

\title{A Taxonomy of Daemons in Self-Stabilization}
\author{Swan Dubois\protect\footnote{UPMC Sorbonne Universit\'es \& INRIA, France, swan.dubois@lip6.fr} \and S\'ebastien Tixeuil\protect\footnote{UPMC Sorbonne Universit\'es \& Institut Universitaire de France, France, sebastien.tixeuil@lip6.fr}}
\date{}

\begin{document}

\maketitle

\begin{abstract}
We survey existing scheduling hypotheses made in the literature in self-stabilization, commonly referred to under the notion of \emph{daemon}. We show that four main characteristics (distribution, fairness, boundedness, and enabledness) are enough to encapsulate the various differences presented in existing work. Our naming scheme makes it easy to compare daemons of particular classes, and to extend existing possibility or impossibility results to new daemons. We further examine existing daemon transformer schemes and provide the exact transformed characteristics of those transformers in our taxonomy.\\
\\

\end{abstract}

\section{Introduction}\label{sec:intro}

\emph{Daemons} are one of the most central yet less understood concepts in self-stabilization. Self-stabilization~\cite{D74j,D00b,T09bc} is a versatile approach to enable forward recovery in distributed systems and networks. Intuitively a distributed system is self-stabilizing if it is able to recover proper behavior after being started from an arbitrary initial global state. This permit to withstand any kind of transient fault or attack (\emph{i.e.} transient in the sense that faults stop occurring after a while) as the recovery mechanism does not make any assumption about what caused the initial arbitrary state.

Self-stabilizing protocols have to fight against two main adversaries that are interdependent. The first adversary is the initial global state. The second adversary is the amount of asynchrony amongst participants. In classical fault-tolerant (\emph{e.g.} crash fault tolerant), more asynchrony usually means more impossibilities~\cite{FLP85j}. In self-stabilization, more synchrony can also be the source of more impossibilities. 

\begin{figure}[htbp]
\begin{center}
\includegraphics[width=0.6\textwidth]{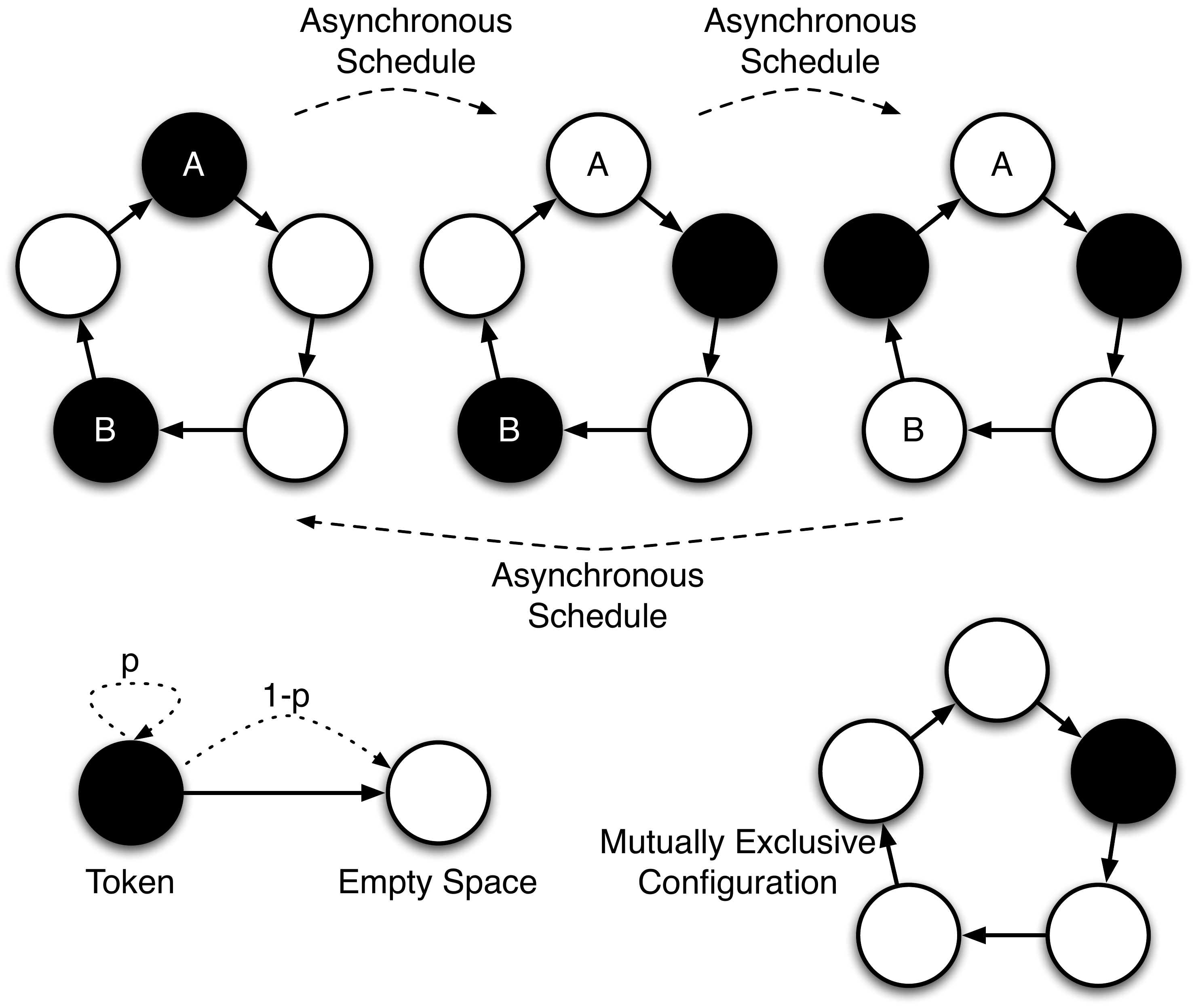}
\end{center}
\caption{Mutual exclusion \emph{vs.} asynchronous scheduling}
\label{fig:mutex}
\end{figure}

Consider for example the mutual exclusion protocol proposed by Herman~\cite{H90j} and depicted in Figure~\ref{fig:mutex}. The protocol operates under the assumption that the network has the shape of a unidirectional ring (processes may only obtain information from their predecessor on the ring, and send information on their successor on the ring). Processes may hold tokens depending on their initial state, and the goal of the protocol is to ensure that regardless of the initial state, the network converges to a point where a single token is present and circulates infinitely often thereafter. Informally, the protocol can be described as follows: whenever a process holds a token, it keeps the token with probability $p$, and sends the token to its immediate successor on the ring with probability $1-p$. If a process holding a token receives a token from its predecessor, the two tokens are merged. This protocol was well studied assuming synchronous scheduling for all processes~\cite{DHT04ca,FMP06j} and convergence to a single token configuration is expected in $\Theta(n^2)$ time units. Now, if process scheduling can be asynchronous, the protocol may not self-stabilize, \emph{i.e.} there may exist an initial state and a particular schedule that prevent tokens from merging. Such an example is presented in Figure~\ref{fig:mutex}: Consider that there exists two initial token in a ring of size five at positions $A$ and $B$. The scheduling is as follows: the process at position $A$ is scheduled for execution until it passes its token (this happens in $O(1)$ expected time), then the process at position $B$ is scheduled for execution until it passes its token (again, this happens in $O(1)$ expected time). The new configuration is isomorphic to the first one, and the schedule repeats. As a result, the two tokens that are initially present never merge, and the protocol does not stabilize.

\begin{figure}[htbp]
\begin{center}
\includegraphics[width=0.5\textwidth]{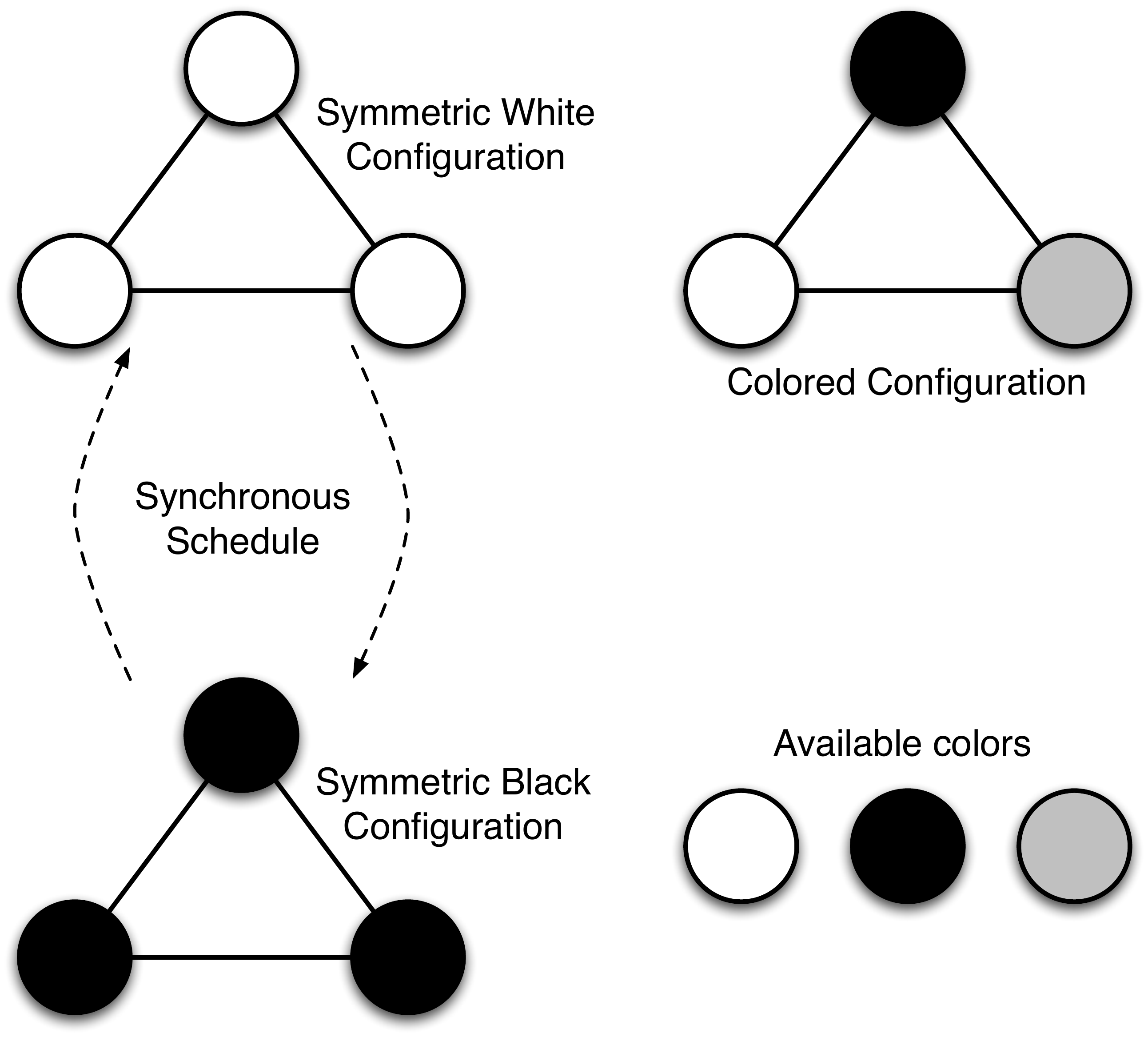}
\end{center}
\caption{Vertex coloring \emph{vs.} synchronous scheduling}
\label{fig:color}
\end{figure}

Another example is the vertex coloring protocol of Gradinariu \emph{et al.}~\cite{GT00c} that is depicted in Figure~\ref{fig:color}. This protocol operates on arbitrary shaped networks under the assumption that no two neighboring processes are scheduled simultaneously. The protocol colors the graph using $deg(g)+1$ colors in a greedy manner, where $deg(g)$ denotes the maximum degree of graph $g$. Whenever a process is scheduled for execution, it checks whether its color conflicts with one of its neighbors (\emph{i.e.} it has the same color as at least one neighbor). If so, it takes the minimal (assuming arbitrary global order on colors) available color to recolor itself. When the scheduling precludes neighbors to be simultaneously activated, the protocol converges to a vertex coloring of the graph. When the scheduling is synchronous, the protocol may not stabilize. Consider the example presented in Figure~\ref{fig:color}: the initial configuration is \emph{symmetric white}, that is, all processes have white color. If all processes are scheduled for execution in this context, they all choose the minimal available color (here, black) and the system reaches a symmetric black configuration. Again, if all processes are scheduled for execution in this context, they all choose the minimal available color (here, white) and the system reaches a symmetric white configuration. The scheduling repeats and the system never stabilizes.

Those two examples are representatives of the assumptions made to ensure stabilization of particular protocols. They also show that depending on the problem to be solved, depending on the protocol used to solve the problem, the class of scheduling hypotheses made is quite different. It is nevertheless appealing yet difficult to relate those two scheduling assumptions in a common framework (one relates to temporal constraints, while the other relates to spatial constraints). Literature presenting self-stabilizing protocols typically abstract scheduling assumptions under the notion of~\emph{daemon}. Intuitively, a daemon is just a predicate on global executions, which could in principle be any possible predicate. If every execution of a particular protocol that satisfies the daemon's predicate converges to a legitimate configuration, the protocol is self-stabilizing under this daemon. 

This approach has the advantage of clearly separating the protocol (that is designed to solve a particular problem) and the scheduling assumptions (that can be seen as an adversary of the protocol, hence the term \emph{daemon}). However, the problem of comparing possibly unrelated daemons may occur \emph{e.g.} when choosing a particular protocol for implementation in a particular environment (\emph{i.e.} assuming a particular daemon). One would generally like to design a protocol for the strongest adversary (that is, the most inclusive defining predicate), while impossibility results should be given for the weakest adversary (that is, the least inclusive defining predicate). Obviously, checking whether a particular solution supports a particular environment (that is, the daemon supported by the solution includes the daemon defining the target environment) or whether a particular problem is solvable in a particular environment (that is, the daemon that makes the problem impossible to solve intersects with the daemon defining the target environment) are important questions a self-stabilizing protocol designer or implementer should be able to answer.

This paper presents a taxonomy for describing daemons having already been used in the self-stabilizing literature. After presenting our model in Section~\ref{sec:model}, we review in Section~\ref{sec:characterization} the four characteristic traits of daemons existing in the literature. In Section~\ref{sec:comparison}, we show how our taxonomy can be used to compare daemons in particular contexts with a ``more powerful'' relation, and maps classical daemons according to their respective power. Section~\ref{sec:transformers} reviews algorithms transformations for turning a daemon into another and depicts the influence of the transformation with respect to all four characteristic daemon traits. Section~\ref{sec:conclusion} provides some concluding remarks.

\section{Model and Definitions}\label{sec:model}
 
\paragraph{Distributed protocol} A distributed system consists of a set of processes that form a communication graph. The processes are vertices in this graph and $V$ denotes the set of vertices. The edges of this graph are pairs of processes that can communicate with each other. Such pairs are neighbors and $E$ denotes the set of edges ($E\subseteq V^2$). Hence, $g=(V,E)$ is the communication graph of the distributed system. Each vertex of $g$ has a set of variables, each of them ranges over a fixed domain of values. A state $\gamma(v)$ of a vertex $v$ is the vector of values of all variables of $v$ at a given time. An assignment of values to all variables of the graph is a configuration. The set of configurations of $g$ is denoted by $\Gamma$. An action $\alpha$ of $g$ transitions the graph from one configuration to another. The set of actions of $g$ is denoted by $A$ ($A=\{(\gamma,\gamma')|\gamma\in\Gamma ,\gamma'\in\Gamma ,\gamma\neq\gamma'\}$). A \emph{distributed protocol} $\pi$ on $g$ is defined as a subset of $A$ that gathers all actions of $g$ allowed by $\pi$. The set of distributed protocols on $g$ is denoted by $\Pi$ ($\Pi=P(A)$\footnote{where, for any set $S$, $P(S)$ denotes the set of parts of $S$.}).

\paragraph{Execution} Given a graph $g$, a distributed protocol $\pi$ on $g$, an \emph{execution} $\sigma$ of $\pi$ on $g$ starting from a given configuration $\gamma_0$ is a maximal sequence of actions of $\pi$ of the following form $\sigma=(\gamma_0,\gamma_1)(\gamma_1,\gamma_2)(\gamma_2,\gamma_3)\ldots$. An execution is \emph{maximal} if it is either infinite or finite but its last configuration is terminal (that is, there exists no actions of $\pi$ starting from this configuration). The set of all executions of $\pi$ on $g$ starting from all configurations of $\Gamma$ is denoted by $\Sigma_\pi$. The set of all executions of all distributed protocols on $S$ starting from all configurations of $\Gamma$ is denoted by $\Sigma_\Pi$ ($\Sigma_\Pi=\{\Sigma_\pi|\pi\in\Pi\}$).

\paragraph{Daemon} The asynchrony of executions is captured by an abstraction called \emph{daemon}. Intuitively, a daemon is a restriction on the executions of distributed protocols to be considered possible. A formal definition follows.

\begin{definition}[Daemon]
Given a graph $g$, a daemon $d$ on $g$ is a function that associates to each distributed protocol $\pi$ on $g$ a subset of executions of $\pi$.
\[\begin{array}{ccrcl}
d & : & \Pi & \longrightarrow & P(\Sigma_\Pi)\\
  &   & \pi & \longmapsto     & d(\pi)\in P(\Sigma_\pi)
\end{array}\]
The set of all daemons on $g$ is denoted by $\mathcal{D}$.
\end{definition}

Given a graph $g$, a daemon $d$ on $g$ and a distributed protocol $\pi$ on $g$, an execution $\sigma$ of $\pi$ ($\sigma\in\Sigma_\pi$) is \emph{allowed} by $d$ if and only if $\sigma\in d(\pi)$. Also, given a graph $g$, a daemon $d$ on $g$ and a distributed protocol $\pi$ on $g$, we say that $\pi$ \emph{runs} on $g$ under $d$ if we consider that only possible executions of $\pi$ on $g$ are those allowed by $d$.

\paragraph{Other Notations} Given a graph $g$ and a distributed protocol $\pi$ on $g$, we introduce the following set of notations. First, $n$ denotes the number of vertices of the graph whereas $m$ denotes the number of edges ($n=|V|$ and $m=|E|$). The distance between two vertices $u$ and $v$ (that is, the length of a shortest path between $u$ and $v$ in $g$) is denoted by $dist(g,u,v)$. The diameter of $g$ (that is, the maximal distance between two vertices of $g$) is denoted by $diam(g)$. The maximal degree of $g$  (that is, the maximal number of neighbors of a vertex in $g$) is denoted by $deg(g)$ (note that $deg^+(g)$ denotes the maximal out-degree when $g$ is oriented). 

Each action of $g$ is characterized by the set of vertices that change their state during the action. We define the following function:
\[\begin{array}{ccrcl}
Act & : & A        & \longrightarrow & P(V)\\
    &   & \alpha=(\gamma,\gamma') & \longmapsto     & \{v\in V|\gamma(v)\neq\gamma'(v)\}
\end{array}\]

A vertex $v$ is enabled by $\pi$ in a configuration $\gamma$ if and only if
\[\exists \gamma'\in\Gamma,(\gamma,\gamma')\in \pi, \gamma(v)\neq\gamma'(v)\]

Each configuration of $g$ is characterized by the set of vertices enabled by $\pi$ in this configuration. We define the following function:
\[\begin{array}{ccrcl}
Ena & : & \Gamma\times\Pi & \longrightarrow & P(V)\\
    &   & (\gamma,\pi)              & \longmapsto     & \{v\in V|v \mbox{ is enabled by } \pi \mbox{ in } \gamma\}
\end{array}\]

\section{Characterization of Daemons}\label{sec:characterization}

In this section, we review the four characteristic traits of daemons existing in the literature, namely \emph{distribution} (Section~\ref{sub:distribution}), \emph{fairness} (Section~\ref{sub:fairness}), \emph{boundedness} (Section~\ref{sub:boundedness}), and \emph{enabledness} (Section~\ref{sub:enabledness}).

\subsection{Distribution}\label{sub:distribution}

Constraints about the spatial scheduling of processes appeared since the seminal paper of Dijkstra~\cite{D74j}, as both the central (a single process is scheduled for execution at any given time) and the distributed (any subset of enabled processes may be scheduled for execution at any given time) daemons are presented. Subsequent literature~\cite{DGT00c,KY02j,DNT09j} enriched the initial model with intermediate steps. Intuitively a daemon is $k$-central is no two processes less than $k$ hops away are allowed to be simultaneously scheduled. A formal definition follows.

\begin{definition}[$k$-Centrality]
Given a graph $g$, a) daemon $d$ is $k$-central if and only if
\[\begin{array}{r@{}l}
\exists k\in\mathbb{N},\forall \pi\in\Pi,\forall \sigma=(\gamma_0,\gamma_1)&(\gamma_1,\gamma_2)\ldots\in d(\pi),\forall i\in\mathbb{N},\forall (u,v)\in V^2,\\
&[u\neq v \wedge u\in Act(\gamma_i,\gamma_{i+1}) \wedge v\in Act(\gamma_i,\gamma_{i+1})]\Rightarrow dist(g,u,v)>k
\end{array}\]
The set of $k$-central daemons is denoted by $k$-$\mathcal{C}$.
\end{definition}

In the literature, a $0$-central daemon is often called \emph{distributed}, and a $diam(g)$-central daemon is either called \emph{central} or \emph{sequential}.

\begin{proposition}\label{prop:distribution}
Given a graph $g$, the following statement holds:
\[\forall k\in\{0,\ldots,diam(g)-1\},(k+1)\mbox{-}\mathcal{C}\subsetneq k\mbox{-}\mathcal{C}\]
\end{proposition}

\begin{proof}
Let $g$ be a graph and $k\in\{0,\ldots,diam(g)-1\}$. We first prove that $(k+1)\mbox{-}\mathcal{C}\subseteq k\mbox{-}\mathcal{C}$.

Let $d$ be a daemon such that $d\in(k+1)\mbox{-}\mathcal{C}$. Then, by definition:
\[\begin{array}{r@{}l}
\forall \pi\in\Pi,\forall \sigma=(\gamma_0,\gamma_1)&(\gamma_1,\gamma_2)\ldots\in d(\pi),\forall i\in\mathbb{N},\forall (u,v)\in V^2,\\
&[u\neq v \wedge u\in Act(\gamma_i,\gamma_{i+1}) \wedge v\in Act(\gamma_i,\gamma_{i+1})]\Rightarrow dist(g,u,v)>k+1
\end{array}\]

As $k<k+1$, we obtain that: $\forall (u,v)\in V^2, dist(g,u,v)>k+1\Rightarrow dist(g,u,v)>k$. As a consequence:
\[\begin{array}{r@{}l}
\forall \pi\in\Pi,\forall \sigma=(\gamma_0,\gamma_1)&(\gamma_1,\gamma_2)\ldots\in d(\pi),\forall i\in\mathbb{N},\forall (u,v)\in V^2,\\
&[u\neq v \wedge u\in Act(\gamma_i,\gamma_{i+1}) \wedge v\in Act(\gamma_i,\gamma_{i+1})]\Rightarrow dist(g,u,v)>k
\end{array}\]

By definition, this implies that $d\in k\mbox{-}\mathcal{C}$ and shows us that $(k+1)\mbox{-}\mathcal{C}\subseteq k\mbox{-}\mathcal{C}$.

There remains to prove that $(k+1)\mbox{-}\mathcal{C}\neq k\mbox{-}\mathcal{C}$. It is sufficient to construct a daemon $d$ such that: $d\in k\mbox{-}\mathcal{C}$ and $d\notin (k+1)\mbox{-}\mathcal{C}$.

Let $d$ be a daemon of $k\mbox{-}\mathcal{C}$ that satisfies:
\[\begin{array}{r@{}l}
\exists \pi\in\Pi, \exists \sigma=(\gamma_0,\gamma_1) & (\gamma_1,\gamma_2)\ldots\in d(\pi), \exists i\in\mathbb{N},\exists (u,v)\in V^2,\\
& u\neq v \wedge u\in Act(\gamma_i,\gamma_{i+1}) \wedge v\in Act(\gamma_i,\gamma_{i+1}) \wedge dist(g,u,v)=k+1 
\end{array}\]

Note that $d$ exists since the execution $\sigma$ is not contradictory with the fact that $d\in k\mbox{-}\mathcal{C}$. On the other hand, we can observe that $d\notin (k+1)\mbox{-}\mathcal{C}$ since the execution $\sigma$ cannot satisfy the definition of an execution allowed by a $(k+1)$-central daemon. This completes the proof of the proposition.
\end{proof}

\begin{figure}[t]
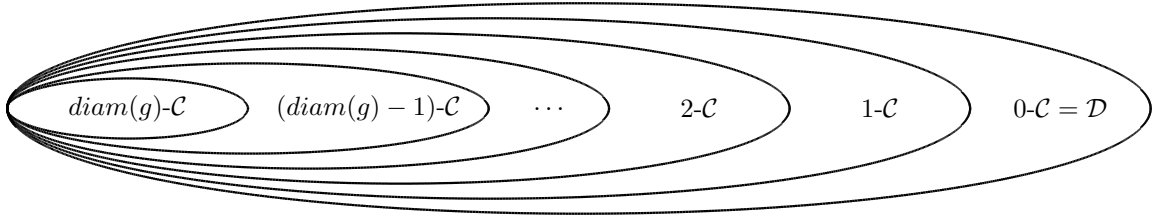

\noindent \begin{centering} \include{lfig/Central}
  \par\end{centering}
 \caption{Inclusions of sets of daemons with respect to distribution.}
\label{fig:central}
\end{figure}

Figure \ref{fig:central} renders Proposition~\ref{prop:distribution} graphically.

\subsection{Fairness}\label{sub:fairness}

The fairness properties of daemons was not discussed in the seminal paper of Dijkstra~\cite{D74j}, as ``executing and action'' was tantamount to ``using critical section'' in its mutual exclusion schemes. So, only global progress was assumed, \emph{i.e.} any set of enabled processes could be scheduled for execution. This very weak assumption was later referred to as an ``unfair'' daemon~\cite{KY97j,DGT00c,KY02j,DGT04j}, since it may happen that a continuously enabled process is never scheduled for execution. In our taxonomy, this ``unfair'' property is simply having no assumptions besides ``distributed''. The notion of \emph{weak fairness}~\cite{K01j,HLCW10j} prevent such behaviors, as it mandates continuously enabled processes to eventually be scheduled by the daemon. A formal definition follows.
 
\begin{definition}[Weak Fairness]
Given a graph $g$, a daemon $d$ is weakly fair if and only if
\[\begin{array}{r@{}l}
\forall \pi\in\Pi,&\forall \sigma=(\gamma_0,\gamma_1)(\gamma_1,\gamma_2)\ldots\in \Sigma_\pi,\\
& [\exists i\in\mathbb{N},\exists v\in V,(\forall j\geq i,v\in Ena(\gamma_j,\pi))\wedge(\forall j\geq i,v\notin Act(\gamma_j,\gamma_{j+1}))]\Rightarrow \sigma\notin d(\pi)
\end{array}\]
A weakly fair daemon is also called a fair daemon. The set of (weakly) fair daemons is denoted by $\mathcal{WF}$ or by $\mathcal{F}$. A daemon that is not fair is called unfair. The set of unfair daemons is denoted by $\bar{\mathcal{F}}$ ($\bar{\mathcal{F}}=\mathcal{D}\setminus\mathcal{F}$).
\end{definition}

For some protocols (including protocols involving Byzantine behaviors~\cite{DPNT10c,DPT11j}), weak fairness is not sufficient to guarantee convergence, and the notion of \emph{strong fairness} was defined~\cite{KC98j,K01j}. Intuitively a daemon is strongly  fair if any process that is enabled infinitely often is eventually scheduled for execution by the daemon. A formal definition follows.

\begin{definition}[Strong Fairness]
Given a graph $g$, a daemon $d$ is strongly fair if and only if
\[\begin{array}{r@{}l}
\forall \pi\in\Pi,&\forall \sigma=(\gamma_0,\gamma_1)(\gamma_1,\gamma_2)\ldots\in \Sigma_\pi,\\
& [\exists i\in\mathbb{N},\exists v\in V,(\forall j\geq i,\exists k\geq j,v\in Ena(\gamma_k,\pi))\wedge(\forall j\geq i,v\notin Act(\gamma_j,\gamma_{j+1}))]\Rightarrow \sigma\notin d(\pi)
\end{array}\]
The set of strongly fair daemons is denoted by $\mathcal{SF}$.
\end{definition}

The strongest notion of fairness (in self-stabilizing systems of finite size) is due to Gouda~\cite{G01cb}. In short, a weakly stabilizing protocol (\emph{i.e.} a protocol such that from any initial configuration, there exists an execution that leads to a legitimate configuration) is in fact self-stabilizing assuming Gouda's notion of fairness. Intuitively, a daemon is \emph{Gouda fair} if from any configuration that appears infinitely often in an execution, every transition is eventually scheduled for execution. A formal definition follows. 

\begin{definition}[Gouda Fairness]
Given a graph $g$, a daemon $d$ is Gouda fair if and only if
\[\begin{array}{r@{}l}
\forall \pi\in\Pi,&\forall \sigma=(\gamma_0,\gamma_1)(\gamma_1,\gamma_2)\ldots\in \Sigma_\pi,\forall (\gamma,\gamma')\in \pi\\
& [\exists i\in\mathbb{N},(\forall j\geq i,\exists k\geq j,\gamma_k=\gamma)\wedge(\forall j\geq i,(\gamma_j,\gamma_{j+1})\neq (\gamma,\gamma'))]\Rightarrow \sigma\notin d(\pi)
\end{array}\]
The set of Gouda fair daemons is denoted by $\mathcal{GF}$.
\end{definition}

\begin{proposition}
\label{prop:fairness}
Given a graph $g$, the following properties hold:
\[\begin{array}{rcl}
\mathcal{GF} & \subsetneq & \mathcal{SF}\\
\mathcal{SF} & \subsetneq & \mathcal{WF}\\
\mathcal{WF} & \subsetneq & \mathcal{D}\\
\end{array}\]
\end{proposition}

\begin{proof}
We first prove that $\mathcal{GF}\subsetneq\mathcal{SF}$. We start by proving that $\mathcal{GF}\subseteq\mathcal{SF}$.

Let $d$ be a daemon of $\mathcal{GF}$. Assume that there exist $\pi\in\Pi$ and $\sigma=(\gamma_0,\gamma_1)(\gamma_1,\gamma_2)\ldots\in \Sigma_\pi$ such that
\[\exists i\in\mathbb{N},\exists v\in V,(\forall j\geq i,\exists k\geq j,v\in Ena(\gamma_k,\pi))\wedge(\forall j\geq i,v\notin Act(\gamma_j,\gamma_{j+1}))\]

Since $\pi$ is a finite subset of actions of $g$, this property implies the following:
\[\exists (\gamma,\gamma')\in \pi, \exists i\in\mathbb{N},(\forall j\geq i,\exists k\geq j,\gamma_k=\gamma)\wedge(\forall j\geq i,(\gamma_j,\gamma_{j+1})\neq (\gamma,\gamma'))\]

As $d\in\mathcal{GF}$, we can deduce that $\sigma\notin d(\pi)$ by definition. Consequently:
\[\begin{array}{r@{}l}
\forall \pi\in\Pi,&\forall \sigma=(\gamma_0,\gamma_1)(\gamma_1,\gamma_2)\ldots\in \Sigma_\pi,\\
& [\exists i\in\mathbb{N},\exists v\in V,(\forall j\geq i,\exists k\geq j,v\in Ena(\gamma_k,\pi))\wedge(\forall j\geq i,v\notin Act(\gamma_j,\gamma_{j+1}))]\Rightarrow \sigma\notin d(\pi)
\end{array}\]

This proves that $d\in\mathcal{SF}$ and hence that $\mathcal{GF}\subseteq\mathcal{SF}$.

It remains to prove that $\mathcal{GF}\neq\mathcal{SF}$. It is sufficient to construct a daemon $d$ such that $d\in\mathcal{SF}$ and $d\notin\mathcal{GF}$.

Let $g$ be a graph and $\pi$ be a distributed protocol such that:
\[\exists (\gamma,\gamma',\gamma'')\in\Gamma^3, (\gamma,\gamma')\in \pi \wedge (\gamma,\gamma'')\in \pi \wedge Act(\gamma,\gamma')=Act(\gamma,\gamma'')\]

Then, it is possible to define a daemon $d\in\mathcal{SF}$ and an execution $\sigma=(\gamma_0,\gamma_1)(\gamma_1,\gamma_2)\ldots\in \Sigma_\pi$ such that:
\[\exists i\in\mathbb{N},(\forall j\geq i,\exists k\geq j,\gamma_k=\gamma)\wedge(\forall j\geq i,\gamma_j=\gamma\Rightarrow(\gamma_j,\gamma_{j+1})=(\gamma,\gamma'')\neq (\gamma,\gamma'))\]

We can conclude that $d\notin\mathcal{GF}$ since the execution $\sigma$ cannot satisfy the definition of an execution allowed by a Gouda fair daemon. That proves the result (since $d\in\mathcal{SF}$ by assumption).

There remains to prove that $\mathcal{SF}\subsetneq\mathcal{WF}$. We first prove that $\mathcal{SF}\subseteq\mathcal{WF}$.

Let $d$ be a daemon of $\mathcal{SF}$. Assume that there exists $\pi\in\Pi$ and $\sigma=(\gamma_0,\gamma_1)(\gamma_1,\gamma_2)\ldots\in \Sigma_\pi$ such that
\[\exists i\in\mathbb{N},\exists v\in V,(\forall j\geq i,v\in Ena(\gamma_j,\pi))\wedge(\forall j\geq i,v\notin Act(\gamma_j,\gamma_{j+1}))\]

This property implies the following:
\[\exists i\in\mathbb{N},\exists v\in V,(\forall j\geq i,\exists k=j, v\in Ena(\gamma_k,\pi))\wedge(\forall j\geq i,v\notin Act(\gamma_j,\gamma_{j+1}))\]

As $d\in\mathcal{SF}$, we can deduce that $\sigma\notin d(\pi)$ by definition. Consequently:
\[\begin{array}{r@{}l}
\forall \pi\in\Pi,&\forall \sigma=(\gamma_0,\gamma_1)(\gamma_1,\gamma_2)\ldots\in \Sigma_\pi,\\
& [\exists i\in\mathbb{N},\exists v\in V,(\forall j\geq i,v\in Ena(\gamma_j,\pi))\wedge(\forall j\geq i,v\notin Act(\gamma_j,\gamma_{j+1}))]\Rightarrow \sigma\notin d(\pi)
\end{array}\]

This proves that $d\in\mathcal{WF}$ and hence that $\mathcal{SF}\subseteq\mathcal{WF}$.

It remains to prove that $\mathcal{SF}\neq\mathcal{WF}$. It is sufficient to construct a daemon $d$ such that $d\in\mathcal{WF}$ and $d\notin\mathcal{SF}$.

Let $g$ be a graph, $\pi$ be a distributed protocol and $u,v$ be two vertices such that:
\[\exists (\gamma,\gamma')\in\Gamma^2,\left\{\begin{array}{l}
v\in Ena(\gamma,\pi) \wedge u\in Ena(\gamma,\pi)\wedge v\notin Ena(\gamma',\pi)\wedge u\in Ena(\gamma',\pi)\\
v\notin Act(\gamma,\gamma') \wedge u\in Act(\gamma,\gamma')\wedge v\notin Act(\gamma',\gamma)\wedge u\in Act(\gamma',\gamma)\\
(\gamma,\gamma')\in \pi \wedge (\gamma',\gamma)\in \pi
\end{array}\right.\]

Then, it is possible to define a daemon $d\in\mathcal{WF}$ and an execution $\sigma=(\gamma_0,\gamma_1)(\gamma_1,\gamma_2)\ldots\in \Sigma_\pi$ such that:
\[\sigma\in d(\pi) \wedge (\forall p\in\mathbb{N},\gamma_{2p}=\gamma\wedge\gamma_{2p+1}=\gamma')\]

We can observe that $\sigma$ satisfies:
\[\exists i=0\in\mathbb{N}, [(\forall j\geq i,(\exists k\geq j, v\in Ena(\gamma_k,\pi))\wedge(\exists k'\geq j, v\notin Ena(\gamma_{k'},\pi))) \wedge (\forall j\geq i,v\notin Act(\gamma_j,\gamma_{j+1}))]\]

We can conclude that $d\notin\mathcal{SF}$ since the execution $\sigma$ cannot satisfy the definition of an execution allowed by a strongly fair daemon. That proves the result (since $d\in\mathcal{WF}$ by assumption).

Finally, we prove that $\mathcal{WF}\subsetneq\mathcal{D}$. As the definition implies that $\mathcal{WF}\subseteq\mathcal{D}$, it remains to prove that $\mathcal{WF}\neq\mathcal{D}$. It is sufficient to construct a daemon $d$ such that $d\in\mathcal{D}$ and $d\notin\mathcal{WF}$.

Let $g$ be a graph and $\pi$ be a distributed protocol such that there exists $v\in V$ satisfying: 
\[\forall (\gamma,\gamma')\in \pi, v\in Ena(\gamma,\pi)\Rightarrow |Ena(\gamma,\pi)|\geq 2\]

Then, it is possible to define a daemon $d$ and an execution $\sigma=(\gamma_0,\gamma_1)(\gamma_1,\gamma_2)\ldots\in \Sigma_\pi$ such that:
\[\forall i\in\mathbb{N},v\notin Act(\gamma_i,\gamma_{i+1})\wedge \sigma\in d(\pi)\]

We can conclude that $d\notin\mathcal{WF}$ since the execution $\sigma$ cannot satisfy the definition of an execution allowed by a weakly fair daemon. That proves the result (since $d\in\mathcal{D}$ by definition).
\end{proof}

\begin{figure}[t]
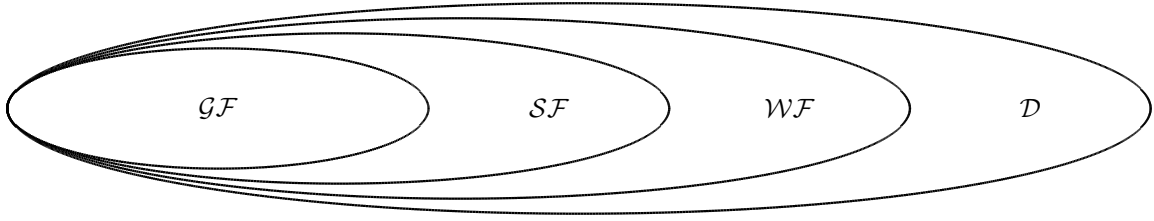

\noindent \begin{centering} \include{lfig/Fairness}
  \par\end{centering}
 \caption{Inclusions of sets of daemons with respect to fairness.}
\label{fig:fairness}
\end{figure}

Figure \ref{fig:fairness} renders Proposition~\ref{prop:fairness} graphically. Devismes \emph{et al.}~\cite{DTY08c} observe that in infinite systems, Gouda fairness is not the strongest form of fairness.

\subsection{Boundedness}\label{sub:boundedness}

Boundedness was first presented in~\cite{DGT00c} as a property achieved by a daemon transformer (see also Section~\ref{sec:transformers}) and was also used as a benchmark to evaluate the performance of self-stabilizing protocols under various kinds of daemons~\cite{BDGM02j,BGJ01c}. Intuitively a daemon is $k$-bounded if no process can be scheduled more than $k$ times between any two schedulings of any other process. Note that this does not imply that there exists a bound on the ``speed'' ratio between any two processes: in particular if a process is never scheduled in a particular execution, another process may be scheduled more than $k$ times in the execution sequel without breaking the $k$-boundedness constraint. As a matter of fact, a daemon can be both $k$-bounded and unfair. A formal definition follows.

\begin{definition}[$k$-Boundedness]
Given a graph $g$, a daemon $d$ is $k$-bounded if and only if
\[\begin{array}{r@{}l}
\exists k\in\mathbb{N}^*,\forall \pi\in\Pi,&\forall \sigma=(\gamma_0,\gamma_1)(\gamma_1,\gamma_2)\ldots\in d(\pi),\forall (i,j)\in\mathbb{N}^2,\forall v\in V,\\
& \big[
[v\in Act(\gamma_i,\gamma_{i+1}) \wedge (\forall l\in\mathbb{N},l<i\Rightarrow v\notin Act(\gamma_l,\gamma_{l+1}))]\\
& \Rightarrow \forall u\in V\setminus\{v\},|\{l\in\mathbb{N}|l< i\wedge u\in Act(\gamma_l,\gamma_{l+1})\}|\leq k
\big]\wedge\\
& \big[
[i<j \wedge v\in Act(\gamma_i,\gamma_{i+1}) \wedge v\in Act(\gamma_j,\gamma_{j+1}) \wedge (\forall l\in\mathbb{N},i<l<j\Rightarrow v\notin Act(\gamma_l,\gamma_{l+1}))]\\
& \Rightarrow \forall u\in V\setminus\{v\},|\{l\in\mathbb{N}|i\leq l< j\wedge u\in Act(\gamma_l,\gamma_{l+1})\}|\leq k
\big]
\end{array}\]
The set of $k$-bounded daemons is denoted by $k$-$\mathcal{B}$. The set of bounded daemons is denoted by $\mathcal{B}$ ($\mathcal{B}=\underset{k\in\mathbb{N}^*}{\bigcup}k\mbox{-}\mathcal{B}$). A daemon that is not $k$-bounded for any $k\in\mathbb{N}^*$ is called unbounded. The set of unbounded daemons is denoted by $\bar{\mathcal{B}}$ ($\bar{\mathcal{B}}=\mathcal{D}\setminus\mathcal{B}$).
\end{definition}

\begin{proposition}\label{prop:boundedness}
Given a graph $g$, the following statements hold:
\[\forall k\in\mathbb{N}^*, 
\left\{
\begin{array}{l} k\mbox{-}\mathcal{B}\subsetneq (k+1)\mbox{-}\mathcal{B}\\
k\mbox{-}\mathcal{B}\subsetneq \mathcal{D}
\end{array}\right.
\]
\end{proposition}

\begin{proof}
Let $g$ be a graph and $k\in\mathbb{N}^*$. We first prove that $k\mbox{-}\mathcal{B}\subseteq (k+1)\mbox{-}\mathcal{B}$.

Let $d$ be a daemon such that $d\in k\mbox{-}\mathcal{B}$. Then, by definition:
\[\begin{array}{r@{}l}
\exists k\in\mathbb{N}^*,\forall \pi\in\Pi,&\forall \sigma=(\gamma_0,\gamma_1)(\gamma_1,\gamma_2)\ldots\in d(\pi),\forall (i,j)\in\mathbb{N}^2,\forall v\in V,\\
& \big[
[v\in Act(\gamma_i,\gamma_{i+1}) \wedge (\forall l\in\mathbb{N},l<i\Rightarrow v\notin Act(\gamma_l,\gamma_{l+1}))]\\
& \Rightarrow \forall u\in V\setminus\{v\},|\{l\in\mathbb{N}|l< i\wedge u\in Act(\gamma_l,\gamma_{l+1})\}|\leq k
\big]\wedge\\
& \big[
[i<j \wedge v\in Act(\gamma_i,\gamma_{i+1}) \wedge v\in Act(\gamma_j,\gamma_{j+1}) \wedge (\forall l\in\mathbb{N},i<l<j\Rightarrow v\notin Act(\gamma_l,\gamma_{l+1}))]\\
& \Rightarrow \forall u\in V\setminus\{v\},|\{l\in\mathbb{N}|i\leq l< j\wedge u\in Act(\gamma_l,\gamma_{l+1})\}|\leq k
\big]
\end{array}\]

As $k<k+1$, we obtain that: 
\[\begin{array}{r@{}l}
\exists k\in\mathbb{N}^*,\forall \pi\in\Pi,&\forall \sigma=(\gamma_0,\gamma_1)(\gamma_1,\gamma_2)\ldots\in d(\pi),\forall (i,j)\in\mathbb{N}^2,\forall v\in V,\\
& \big[
[v\in Act(\gamma_i,\gamma_{i+1}) \wedge (\forall l\in\mathbb{N},l<i\Rightarrow v\notin Act(\gamma_l,\gamma_{l+1}))]\\
& \Rightarrow \forall u\in V\setminus\{v\},|\{l\in\mathbb{N}|l< i\wedge u\in Act(\gamma_l,\gamma_{l+1})\}|\leq k+1
\big]\wedge\\
& \big[
[i<j \wedge v\in Act(\gamma_i,\gamma_{i+1}) \wedge v\in Act(\gamma_j,\gamma_{j+1}) \wedge (\forall l\in\mathbb{N},i<l<j\Rightarrow v\notin Act(\gamma_l,\gamma_{l+1}))]\\
& \Rightarrow \forall u\in V\setminus\{v\},|\{l\in\mathbb{N}|i\leq l< j\wedge u\in Act(\gamma_l,\gamma_{l+1})\}|\leq k+1
\big]
\end{array}\]

By definition, this implies that $d\in (k+1)\mbox{-}\mathcal{B}$ and shows us that $k\mbox{-}\mathcal{B}\subseteq (k+1)\mbox{-}\mathcal{B}$.

There remains to prove that $k\mbox{-}\mathcal{B}\neq (k+1)\mbox{-}\mathcal{B}$. It is sufficient to construct a daemon $d$ such that: $d\in (k+1)\mbox{-}\mathcal{B}$ and $d\notin k\mbox{-}\mathcal{B}$.

Let $d$ be a daemon of $(k+1)\mbox{-}\mathcal{B}$ that satisfies:
\[\begin{array}{r@{}l}
\exists \pi\in\Pi, \exists \sigma= & (\gamma_0,\gamma_1)(\gamma_1,\gamma_2)\ldots\in d(\pi), \exists (i,j)\in\mathbb{N}^2,\exists v\in V,\\
& i<j \wedge v\in Act(\gamma_i,\gamma_{i+1}) \wedge v\in Act(\gamma_j,\gamma_{j+1}) \wedge (\forall l\in\mathbb{N},i<l<j\Rightarrow v\notin Act(\gamma_l,\gamma_{l+1}))\\
& \wedge (\exists u\in V\setminus\{v\},|\{l\in\mathbb{N}|i\leq l<j\wedge u\in Act(\gamma_l,\gamma_{l+1})\}|= k+1)
\end{array}\]

Note that $d$ exists since the execution $\sigma$ is not contradictory with the fact that $d\in (k+1)\mbox{-}\mathcal{B}$. On the other hand, we can observe that $d\notin k\mbox{-}\mathcal{B}$ since the execution $\sigma$ cannot satisfy the definition of an execution allowed by a $k$-bounded daemon. This completes the proof of the first property.

Finally, we prove that $k\mbox{-}\mathcal{B}\subsetneq \mathcal{D}$. By definition, $k\mbox{-}\mathcal{B}\subseteq \mathcal{D}$. There remains to prove that $k\mbox{-}\mathcal{B}\neq \mathcal{D}$. By the first property, there exists a daemon $d$ such that $d\in (k+1)\mbox{-}\mathcal{B}$ and $d\notin k\mbox{-}\mathcal{B}$. By definition, $(k+1)\mbox{-}\mathcal{B}\subseteq \mathcal{D}$ holds. Hence, the claimed result.
\end{proof}

\begin{figure}[t]
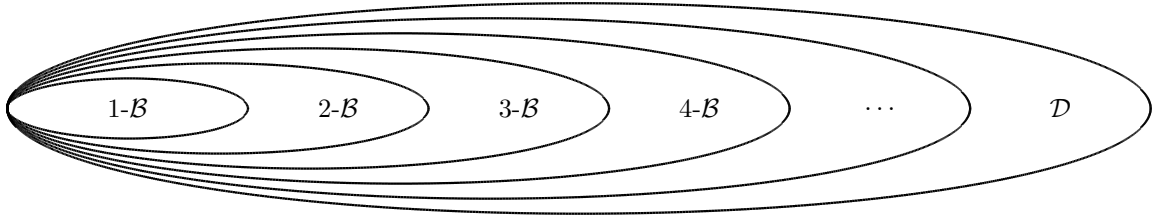

\noindent \begin{centering} \include{lfig/Bounded}
  \par\end{centering}
 \caption{Inclusions of sets of daemons with respect to boundedness.}
\label{fig:bounded}
\end{figure}

Figure \ref{fig:bounded} renders Proposition~\ref{prop:boundedness} graphically.

\subsection{Enabledness}\label{sub:enabledness}

Enabledness is a characterization of daemon properties that is introduced in this paper. It is defined to be related to the intuitive notion that the ratio between the ``speed'' of the fastest process and that of the slowest process is bounded. In an asynchronous setting where we use configurations and time-independent transitions between configurations, $k$-enabledness intuitively means that a particular process can not be enabled more than $k$ times before being activated. A formal definition follows. 

\begin{definition}[$k$-Enabledness]
Given a graph $g$, a daemon $d$ is $k$-enabled if and only if
\[\begin{array}{r@{}l}
\exists k\in\mathbb{N},\forall \pi\in\Pi,&\forall \sigma=(\gamma_0,\gamma_1)(\gamma_1,\gamma_2)\ldots\in d(\pi),\forall (i,j)\in\mathbb{N}^2,\forall v\in V,\\
& \big[
[v\in Act(\gamma_i,\gamma_{i+1})\wedge(\forall l\in\mathbb{N},l<i\Rightarrow v\notin Act(\gamma_l,\gamma_{l+1}))]\\
& \Rightarrow |\{l\in\mathbb{N}|l<i\wedge v\in Ena(\gamma_l,\pi)\}|\leq k
\big]\wedge\\
& \big[
[i<j \wedge v\in Act(\gamma_i,\gamma_{i+1}) \wedge v\in Act(\gamma_j,\gamma_{j+1}) \wedge (\forall l\in\mathbb{N},i<l<j\Rightarrow v\notin Act(\gamma_l,\gamma_{l+1}))]\\
& \Rightarrow |\{l\in\mathbb{N}|i<l<j\wedge v\in Ena(\gamma_l,\pi)\}|\leq k
\big]\wedge\\
& \big[
[v\in Act(\gamma_i,\gamma_{i+1})\wedge(\forall l\in\mathbb{N},l>i\Rightarrow v\notin Act(\gamma_l,\gamma_{l+1}))]\\
& \Rightarrow |\{l\in\mathbb{N}|l>i\wedge v\in Ena(\gamma_l,\pi)\}|\leq 
k\big]
\end{array}\]
The set of $k$-enabled daemons is denoted by $k$-$\mathcal{E}$.  The set of daemons of bounded enabledness is denoted by $\mathcal{E}$ ($\mathcal{E}=\underset{k\in\mathbb{N}}{\bigcup}k\mbox{-}\mathcal{E}$). A daemon that is not $k$-enabled for any $k\in\mathbb{N}$ has an unbounded enabledness. The set of daemons of unbounded enabledness is denoted by $\bar{\mathcal{E}}$ ($\bar{\mathcal{E}}=\mathcal{D}\setminus\mathcal{E}$).
\end{definition}

\begin{proposition}
\label{prop:enabledness}
Given a graph $g$, the following statements hold:
\[\forall k\in\mathbb{N}, 
\left\{
\begin{array}{l} k\mbox{-}\mathcal{E}\subsetneq (k+1)\mbox{-}\mathcal{E}\\
k\mbox{-}\mathcal{E}\subsetneq \mathcal{D}
\end{array}\right.
\]
\end{proposition}

\begin{proof}
Let $g$ be a graph and $k\in\mathbb{N}$. We first prove that $k\mbox{-}\mathcal{E}\subseteq (k+1)\mbox{-}\mathcal{E}$.

Let $d$ be a daemon such that $d\in k\mbox{-}\mathcal{E}$. Then, by definition:
\[\begin{array}{r@{}l}
\exists k\in\mathbb{N},\forall \pi\in\Pi,&\forall \sigma=(\gamma_0,\gamma_1)(\gamma_1,\gamma_2)\ldots\in d(\pi),\forall (i,j)\in\mathbb{N}^2,\forall v\in V,\\
& \big[
[v\in Act(\gamma_i,\gamma_{i+1})\wedge(\forall l\in\mathbb{N},l<i\Rightarrow v\notin Act(\gamma_l,\gamma_{l+1}))]\\
& \Rightarrow |\{l\in\mathbb{N}|l<i\wedge v\in Ena(\gamma_l,\pi)\}|\leq k
\big]\wedge\\
& \big[
[i<j \wedge v\in Act(\gamma_i,\gamma_{i+1}) \wedge v\in Act(\gamma_j,\gamma_{j+1}) \wedge (\forall l\in\mathbb{N},i<l<j\Rightarrow v\notin Act(\gamma_l,\gamma_{l+1}))]\\
& \Rightarrow |\{l\in\mathbb{N}|i<l<j\wedge v\in Ena(\gamma_l,\pi)\}|\leq k
\big]\wedge\\
& \big[
[v\in Act(\gamma_i,\gamma_{i+1})\wedge(\forall l\in\mathbb{N},l>i\Rightarrow v\notin Act(\gamma_l,\gamma_{l+1}))]\\
& \Rightarrow |\{l\in\mathbb{N}|l>i\wedge v\in Ena(\gamma_l,\pi)\}|\leq 
k\big]
\end{array}\]

As $k<k+1$, we obtain that: 
\[\begin{array}{r@{}l}
\exists k\in\mathbb{N},\forall \pi\in\Pi,&\forall \sigma=(\gamma_0,\gamma_1)(\gamma_1,\gamma_2)\ldots\in d(\pi),\forall (i,j)\in\mathbb{N}^2,\forall v\in V,\\
& \big[
[v\in Act(\gamma_i,\gamma_{i+1})\wedge(\forall l\in\mathbb{N},l<i\Rightarrow v\notin Act(\gamma_l,\gamma_{l+1}))]\\
& \Rightarrow |\{l\in\mathbb{N}|l<i\wedge v\in Ena(\gamma_l,\pi)\}|\leq k+1
\big]\wedge\\
& \big[
[i<j \wedge v\in Act(\gamma_i,\gamma_{i+1}) \wedge v\in Act(\gamma_j,\gamma_{j+1}) \wedge (\forall l\in\mathbb{N},i<l<j\Rightarrow v\notin Act(\gamma_l,\gamma_{l+1}))]\\
& \Rightarrow |\{l\in\mathbb{N}|i<l<j\wedge v\in Ena(\gamma_l,\pi)\}|\leq k+1
\big]\wedge\\
& \big[
[v\in Act(\gamma_i,\gamma_{i+1})\wedge(\forall l\in\mathbb{N},l>i\Rightarrow v\notin Act(\gamma_l,\gamma_{l+1}))]\\
& \Rightarrow |\{l\in\mathbb{N}|l>i\wedge v\in Ena(\gamma_l,\pi)\}|\leq 
k+1\big]
\end{array}\]

By definition, this implies that $d\in (k+1)\mbox{-}\mathcal{E}$ and shows us that $k\mbox{-}\mathcal{E}\subseteq (k+1)\mbox{-}\mathcal{E}$.

There remains to prove that $k\mbox{-}\mathcal{E}\neq (k+1)\mbox{-}\mathcal{E}$. It is sufficient to construct a daemon $d$ such that: $d\in (k+1)\mbox{-}\mathcal{E}$ and $d\notin k\mbox{-}\mathcal{E}$.

Let $d$ be a daemon of $(k+1)\mbox{-}\mathcal{E}$ that satisfies:
\[\begin{array}{r@{}l}
\exists \pi\in\Pi, \exists \sigma= & (\gamma_0,\gamma_1)(\gamma_1,\gamma_2)\ldots\in d(\pi),\exists (i,j)\in\mathbb{N}^2,\exists v\in V,\\
& i<j \wedge v\in Act(\gamma_i,\gamma_{i+1}) \wedge v\in Act(\gamma_j,\gamma_{j+1}) \wedge (\forall l\in\mathbb{N},i<l<j\Rightarrow v\notin Act(\gamma_l,\gamma_{l+1}))\\
& \wedge |\{l\in\mathbb{N}|i<l<j\wedge v\in Ena(\gamma_l,\pi)\}|=k+1
\end{array}\]

Note that $d$ exists since the execution $\sigma$ is not contradictory with the fact that $d\in (k+1)\mbox{-}\mathcal{E}$. On the other hand, we can observe that $d\notin k\mbox{-}\mathcal{E}$ since the execution $\sigma$ cannot satisfy the definition of an execution allowed by a $k$-enabled daemon. This completes the proof of the first property.

Finally, we prove that $k\mbox{-}\mathcal{E}\subsetneq \mathcal{D}$. By definition, $k\mbox{-}\mathcal{E}\subseteq \mathcal{D}$ holds. There remains to prove that $k\mbox{-}\mathcal{E}\neq \mathcal{D}$. By the first property, there exists a daemon $d$ such that $d\in (k+1)\mbox{-}\mathcal{E}$ and $d\notin k\mbox{-}\mathcal{E}$. By definition, $(k+1)\mbox{-}\mathcal{E}\subseteq \mathcal{D}$ holds. Hence the claimed result.
\end{proof}

\begin{figure}[t]
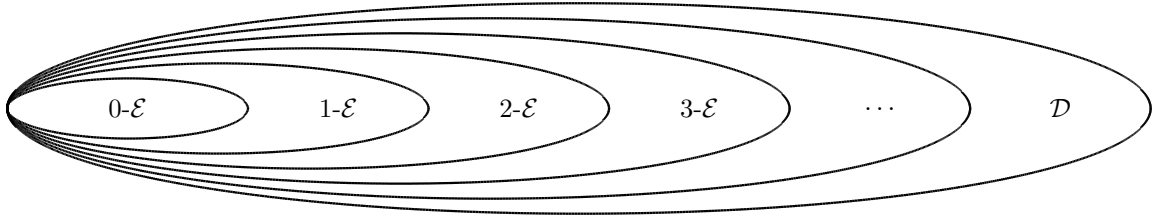

\noindent \begin{centering} \include{lfig/Enabled}
  \par\end{centering}
 \caption{Inclusions of sets of daemons with respect to enabledness.}
\label{fig:enabled}
\end{figure}

Figure \ref{fig:enabled} renders Proposition~\ref{prop:enabledness} graphically. Unlike previous characteristic properties of daemons, enabledness is not completely independent from others. Relationship between enabledness and fairness and boundedness are depicted in the sequel (Sections~\ref{sec:fairenable} and \ref{sec:boundedenable}).

\subsubsection{Relationship between Fairness and Enabledness}\label{sec:fairenable}

Daemons with bounded enabledness cannot ignore scheduling processes more than $k$ times, implying that the overall schedule is at least weakly fair. Nevertheless, the following proposition shows that the converse is not true (\emph{i.e.} there exist daemons that are weakly fair but do not have bounded enabledness, furthermore those daemons are not strongly fair either). There also exist daemons that are strongly fair or Gouda fair, yet do not have finite enabledness.

\begin{proposition}
For any given graph $g$, the following statements hold:
\[\begin{array}{c}
\forall d\in\mathcal{D}, d\in\mathcal{E}\Rightarrow d\in \mathcal{WF}\\
\exists d\in\mathcal{WF}\setminus(\mathcal{E}\cup\mathcal{SF})\\
\exists d\in\mathcal{SF}\setminus(\mathcal{E}\cup\mathcal{GF})\\
\exists d\in\mathcal{GF}\setminus\mathcal{E}
\end{array}\]
\end{proposition}

\begin{proof}
Let $g$ be a graph. Let $d$ be a daemon such that $d\in\mathcal{E}$. Then, there exists $k\in\mathbb{N}$ such that $d\in k\mbox{-}\mathcal{E}$. We are going to prove that $d\in\mathcal{WF}$.

Assume that $\pi$ is a distributed protocol and $\sigma=(\gamma_0,\gamma_1)(\gamma_1,\gamma_2)\ldots$ is an execution of $d(\pi)$ satisfying:
\[\exists i\in\mathbb{N}^*,\exists v\in V, v\in Act(\gamma_{i-1},\gamma_i)\wedge (\forall j\geq i,v\in Ena(\gamma_j,\pi))\wedge(\forall j\geq i,v\notin Act(\gamma_j,\gamma_{j+1}))\]

Then, we obtain:
\[[v\in Act(\gamma_{i-1},\gamma_i)\wedge(\forall l\in\mathbb{N},l>i\Rightarrow v\notin Act(\gamma_l,\gamma_{l+1}))]\wedge |\{l\in\mathbb{N}|l>i\wedge v\in Ena(\gamma_l,\pi)\}|=\infty> k\]

This property is contradictory with $\sigma\in d(\pi)$ and $d\in k\mbox{-}\mathcal{E}$. hence, we deduct that:
\[\begin{array}{r@{}l}
\forall \pi\in\Pi,&\forall \sigma=(\gamma_0,\gamma_1)(\gamma_1,\gamma_2)\ldots\in \Sigma_\pi,\\
& [\exists i\in\mathbb{N},\exists v\in V,(\forall j\geq i,v\in Ena(\gamma_j,\pi))\wedge(\forall j\geq i,v\notin Act(\gamma_j,\gamma_{j+1}))]\Rightarrow \sigma\notin d(\pi)
\end{array}\]

That means that $d\in\mathcal{WF}$ and then, we proved that: $\forall d\in\mathcal{D},d\in\mathcal{E}\Rightarrow d\in \mathcal{WF}$.

There remains to prove that $\exists d\in\mathcal{GF}\setminus\mathcal{E}$. Consider a daemon $d$ such that $d\in\mathcal{GF}$ and a distributed protocol $\pi_1$ such that:
\[\exists (\gamma_0,\gamma_1,\gamma_2)\in\Gamma^3,\exists v\in V,
\left\{\begin{array}{l}
(\gamma_0,\gamma_1)\in \pi_1\wedge v\in Ena(\gamma_0,\pi_1) \wedge v\in Act(\gamma_0,\gamma_1)\\
(\gamma_1,\gamma_2)\in \pi_1\wedge v\in Ena(\gamma_1,\pi_1) \wedge v\notin Act(\gamma_1,\gamma_2)\\
(\gamma_2,\gamma_1)\in \pi_1\wedge v\in Ena(\gamma_2,\pi_1) \wedge v\notin Act(\gamma_2,\gamma_1)
\end{array}\right.\]

Let $\sigma$ be an execution of $d(\pi_1)$ starting from $\gamma_2$. Now, we define the following set of executions of $\pi_1$ (where the product operator denotes the concatenation of portions of executions):
\[\forall k\in\mathbb{N}, \sigma_k=(\gamma_0,\gamma_1)(\gamma_1,\gamma_2).\big[(\gamma_2,\gamma_1)(\gamma_1,\gamma_2)\big]^k.\sigma\]

We can define a daemon $d'$ in the following way:
\[\left\{\begin{array}{l}
\forall \pi\in\Pi\setminus\{\pi_1\},d'(\pi)=d(\pi)\\
d'(\pi_1)=d(\pi_1)\cup\{\sigma_k|k\in\mathbb{N}\}
\end{array}\right.\]

Then, we can observe that $d'\in\mathcal{GF}$ by construction and that, for any $k\in\mathbb{N}$, the execution $\sigma_k\in d'(\pi_1)$ does not satisfy the definition of $k$-enabledness. Consequently, we prove that: $d'\in\mathcal{GF}\setminus\mathcal{E}$. If we follow the same reasoning starting from a daemon $d$ in $\mathcal{WF}\setminus\mathcal{SF}$ (respectively in $\mathcal{SF}\setminus\mathcal{GF}$), we prove that $d'\in\mathcal{WF}\setminus(\mathcal{E}\cup\mathcal{SF})$ (respectively that $d'\in\mathcal{SF}\setminus(\mathcal{E}\cup\mathcal{GF})$), which ends the proof.
\end{proof}

\subsubsection{Relationship between Boundedness and Enabledness}
\label{sec:boundedenable}

As previously mentioned, there is not relationship between boundedness and fairness. In this section, we prove that there is a connexion between (finite) enabledness and (finite) boundedness. In particular, if a daemon is both $k$-enabled and $k'$-bounded (for some particular integers $k$ and $k'$), then $k\leq (n-1)\times k'$ (where $n$ denotes the number of processes in the system). However, there exist daemons that are $k$-enabled (for some integer $k$) but do not have finite boundedness, and daemons that are $k'$-bounded (for some integer $k'$) but do not have finite enabledness. 

\begin{proposition}
For any given graph $g$, the following statements hold:
\[\begin{array}{c}
\forall d\in\mathcal{D},\forall (k,k')\in\mathbb{N}\times\mathbb{N}^*, (d\in k\mbox{-}\mathcal{E}\wedge d\in k'\mbox{-}\mathcal{B})\Rightarrow k\leq (n-1)\times k'\\
\forall k\in\mathbb{N},\exists d\in k\mbox{-}\mathcal{E}\setminus\mathcal{B}\\
\forall k\in\mathbb{N}^*,\exists d\in k\mbox{-}\mathcal{B}\setminus\mathcal{E}
\end{array}\]
\end{proposition}

\begin{proof}
Firstly, we prove that $\forall d\in\mathcal{D},\forall (k,k')\in\mathbb{N}\times\mathbb{N}^*, (d\in k\mbox{-}\mathcal{E}\wedge d\in k'\mbox{-}\mathcal{B})\Rightarrow k\leq (n-1)\times k'$. Consider a daemon $d$ such that $d\in k\mbox{-}\mathcal{E}$ and $d\in k'\mbox{-}\mathcal{B}$ for two given $(k,k')\in\mathbb{N}\times\mathbb{N}^*$.

As $d$ is $k'$-bounded, between two consecutive actions of any vertex $v$, any vertex $u$ such that $u\neq v$ takes at most $k'$ actions. This implies that there exists at most $(n-1)\times k'$ actions between two consecutive actions of $v$ (since the daemon must ensure the progress). This implies that, between two consecutive actions of $v$, there exists at most $(n-1)\times k'$ configurations where $v$ is enabled (without being activated by construction). As $d$ has a bounded enabledness $k$, we can deduce that $k\leq (n-1)\times k'$, which proves the result.

Secondly, we prove that $\forall k\in\mathbb{N},\exists d\in k\mbox{-}\mathcal{E}\setminus\mathcal{B}$. Consider $k\in\mathbb{N}$, a daemon $d$ such that $d\in k\mbox{-}\mathcal{E}$ and a distributed protocol $\pi_1$ such that:
\[\forall\ell\in\mathbb{N}^*,\exists (\gamma_{\ell+1},\gamma_{\ell},\ldots,\gamma_1,\gamma_0)\in\Gamma^{\ell+2},\exists v\in V,
\left\{\begin{array}{l}
\forall i\in\{0,\ldots,\ell\},(\gamma_{i+1},\gamma_i)\in \pi_1\\
\forall i\in\{0,\ldots,\ell\},Act(\gamma_{i+1},\gamma_i)=Ena(\gamma_{i+1},\pi_1)=V
\end{array}\right.\]

Let $\sigma$ be an execution of $d(\pi_1)$ starting from $\gamma_0$. Now, we define the following set of executions of $\pi_1$ (where the product operator denotes the concatenation of portions of executions):
\[\forall k'\in\mathbb{N}^*, \sigma_{k'}=(\gamma_{k'+1},\gamma_{k'})(\gamma_{k'},\gamma_{k'-1})\ldots(\gamma_2,\gamma_1)(\gamma_1,\gamma_0).\sigma\]

Note that, for any $k'\in\mathbb{N}^*$, the portion of execution $(\gamma_{k'+1},\gamma_{k'})(\gamma_{k'},\gamma_{k'-1})\ldots(\gamma_2,\gamma_1)(\gamma_1,\gamma_0)$ is $0$-enabled. Hence, any execution of $\{\sigma_{k'}|k'\in\mathbb{N}^*\}$ is $k$-enabled.

We can define a daemon $d'$ in the following way:
\[\left\{\begin{array}{l}
\forall \pi\in\Pi\setminus\{\pi_1\},d'(\pi)=d(\pi)\\
d'(\pi_1)=d(\pi_1)\cup\{\sigma_{k'}|k'\in\mathbb{N}^*\}
\end{array}\right.\]

Then, we can observe that $d'\in k\mbox{-}\mathcal{E}$ by construction and that, for any $k'\in\mathbb{N}^*$, the execution $\sigma_{k'}\in d'(\pi_1)$ does not satisfy the definition of $k'$-boundedness. Consequently, we prove that: $d'\in k\mbox{-}\mathcal{E}\setminus\underset{k'\in\mathbb{N}^*}{\bigcup}k'\mbox{-}\mathcal{B}=k\mbox{-}\mathcal{E}\setminus\mathcal{B}$. 

Finally, we prove that $\forall k\in\mathbb{N}^*,\exists d\in k\mbox{-}\mathcal{B}\setminus\mathcal{E}$. Consider $k\in\mathbb{N}^*$, a daemon $d$ such that $d\in k\mbox{-}\mathcal{B}$ and a distributed protocol $\pi_1$ such that:
\[\exists (\gamma_0,\gamma_1,\gamma_2)\in\Gamma^3,\exists v\in V,
\left\{\begin{array}{l}
(\gamma_0,\gamma_1)\in \pi_1\wedge v\in Ena(\gamma_0,\pi_1) \wedge Act(\gamma_0,\gamma_1)=\{v\}\\
(\gamma_1,\gamma_2)\in \pi_1\wedge v\in Ena(\gamma_1,\pi_1) \wedge v\notin Act(\gamma_1,\gamma_2)\\
(\gamma_2,\gamma_1)\in \pi_1\wedge v\in Ena(\gamma_2,\pi_1) \wedge v\notin Act(\gamma_2,\gamma_1)\\
Act(\gamma_1,\gamma_2)=Act(\gamma_2,\gamma_1)
\end{array}\right.\]

Let $\sigma$ be an execution of $d(\pi_1)$ starting from $\gamma_2$. Now, we define the following set of executions of $\pi_1$ (where the product operator denotes the concatenation of portions of executions):
\[\forall k'\in\mathbb{N}, \sigma_{k'}=(\gamma_0,\gamma_1)(\gamma_1,\gamma_2).\big[(\gamma_2,\gamma_1)(\gamma_1,\gamma_2)\big]^{k'}.e\]

Note that, for any $k'\in\mathbb{N}$, the portion of execution $(\gamma_0,\gamma_1)(\gamma_1,\gamma_2).\big[(\gamma_2,\gamma_1)(\gamma_1,\gamma_2)\big]^{k'}$ is $1$-bounded. Hence, any execution of $\{\sigma_{k'}|k'\in\mathbb{N}\}$ is $k$-bounded.

We can define a daemon $d'$ in the following way:
\[\left\{\begin{array}{l}
\forall \pi\in\Pi\setminus\{\pi_1\},d'(\pi)=d(\pi)\\
d'(\pi_1)=d(\pi_1)\cup\{\sigma_{k'}|k'\in\mathbb{N}\}
\end{array}\right.\]

Then, we can observe that $d'\in k\mbox{-}\mathcal{B}$ by construction and that, for any $k'\in\mathbb{N}$, the execution $\sigma_{k'}\in d'(\pi_1)$ does not satisfy the definition of $k'$-enabledness. Consequently, we prove that: $d'\in k\mbox{-}\mathcal{B}\setminus\underset{k'\in\mathbb{N}}{\bigcup}k'\mbox{-}\mathcal{E}=k\mbox{-}\mathcal{B}\setminus\mathcal{E}$. 
\end{proof}

\section{Comparing Daemons}\label{sec:comparison}

The four main characteristics presented in Section~\ref{sec:characterization} provide a convenient way to define a particular class of daemons: this class simply combines the four characteristic properties. A formal definition follows.

\begin{definition}[Daemon class]
Given a graph $g$ and four sets of daemons 
\[\left\{\begin{array}{rcll}
C & \in & \{k\mbox{-}\mathcal{C}|k\in\{0,\ldots,diam(g)\}\} & \\
B & \in & \{\mathcal{D},k\mbox{-}\mathcal{B}|k\in\mathbb{N}^*\} & ,\\
E & \in & \{\mathcal{D},k\mbox{-}\mathcal{E}|k\in\mathbb{N}\} & \\
F & \in & \{\mathcal{D},\mathcal{WF},\mathcal{SF},\mathcal{GF}\} &
\end{array}\right.\]
the \emph{class} of daemons $\mathcal{D}(C,B,E,F)$ is defined by $\mathcal{D}(C,B,E,F)=C\cap B\cap E \cap F$.
\end{definition}

\subsection{Comparing daemon classes}

Now, each particular daemon instance $d$ may belongs to several classes (those that include all possible executions under $d$). It is convenient to refer to the \emph{minimal class} of $d$ as the set of characteristics that strictly define $d$. A formal definition follows.

\begin{definition}[Minimal class]
Given a graph $g$ and a daemon $d$, the \emph{minimal class} of $d$ is the class of daemons $\mathcal{D}(C,B,E,F)$ such that:
\[\left\{\begin{array}{l}
d \in \mathcal{D}(C,B,E,F)\\
\forall \mathcal{D}(C',B',E',F')\subsetneq \mathcal{D}(C,B,E,F), d \notin \mathcal{D}(C',B',E',F')
\end{array}\right.\]
\end{definition}

In any particular class, the canonical daemon of this class is a representative element of that class such that for any daemon $d$ in the class, any execution allowed by $d$ is also allowed by the canonical daemon. Simply put, the canonical daemon of a class is the largest element of this class with respect to allowed executions. A formal definition follows.

\begin{definition}[Canonical Daemon]
For a given graph $g$ and a class of daemons $\mathcal{D}(C,B,E,F)$, the canonical daemon $d(C,B,E,F)$ of $\mathcal{D}(C,B,E,F)$ is the daemon defined by:
\[\left\{\begin{array}{l}
d(C,B,E,F)\in\mathcal{D}(C,B,E,F)\\
\forall d\in\mathcal{D}(C,B,E,F),\forall \pi\in\Pi,\forall \sigma\in\Sigma_\pi, \sigma\in d(\pi)\Rightarrow \sigma\in d(C,B,E,F)(\pi)
\end{array}\right.\]
\end{definition}

This way of viewing daemons as a set of possible executions (for a particular graph $g$) drives a natural ``more powerful'' relation definition. For a particular graph $g$, a daemon $d$ is more powerful than another daemon $d'$ if all executions allowed by $d'$ are also allowed by $d$. Overall, $d$ has more scheduling choices than $d'$. A formal definition follows.

\begin{definition}[More powerful relation]
For a given graph $g$, we define the following binary relation $\preccurlyeq$ on $\mathcal{D}$:
\[\forall (d,d')\in\mathcal{D},d\preccurlyeq d'\Leftrightarrow (\forall \pi\in\Pi,d(\pi)\subseteq d'(\pi))\]
If two daemons $d$ and $d'$ satisfy $d\preccurlyeq d'$, we say that $d'$ is more powerful than $d$.
\end{definition}

As with set inclusions, this ``more powerful'' relation induces a partial order, which is formally presented in the sequel.

\begin{proposition}
For any graph $g$, the binary relation $\preccurlyeq$ is a partial order on $\mathcal{D}$.
\end{proposition}

\begin{proof}
Let $g$ be a graph. We are going to prove that the binary relation $\preccurlyeq$ is reflexive, antisymmetric and transitive. Then we show that this order is not total (\emph{i.e.} that there exists some incomparable elements by $\preccurlyeq$ in $\mathcal{D}$).

For any daemon $d\in\mathcal{D}$, we have $\forall\pi\in\Pi,d(\pi)\subseteq d(\pi)$, which proves that $\forall d\in\mathcal{D},d\preccurlyeq d$ (reflexivity of the binary relation $\preccurlyeq$).

Let $d$ and $d'$ be two daemons such that $d\preccurlyeq d'$ and $d'\preccurlyeq d$. Then, by definition:
\[\left.\begin{array}{c}
\forall \pi\in\Pi,d(\pi)\subseteq d'(\pi)\\
\forall \pi\in\Pi,d'(\pi)\subseteq d(\pi)
\end{array}\right\}\Rightarrow \forall \pi\in\Pi,d(\pi)=d'(\pi)\]
In other words, $d=d'$ (antisymmetry of the binary relation $\preccurlyeq$).

Let $d$, $d'$ and $d''$ be three daemons such that $d\preccurlyeq d'$ and $d'\preccurlyeq d''$. Then, by definition:
\[\left.\begin{array}{c}
\forall \pi\in\Pi,d(\pi)\subseteq d'(\pi)\\
\forall \pi\in\Pi,d'(\pi)\subseteq d''(\pi)
\end{array}\right\}\Rightarrow \forall \pi\in\Pi,d(\pi)\subseteq d''(\pi)\]
In other words, $d\preccurlyeq d''$ (transitivity of the binary relation $\preccurlyeq$).

Let $d$ be a daemon, $\pi_1$ and $\pi_2$ be two distributed protocols and $\sigma_1$ and $\sigma_2$ be two executions such that:
\[\left\{\begin{array}{c}
\pi_1\neq\pi_2\\
\sigma_1\notin d(\pi_1)\\
\sigma_2\notin d(\pi_2)
\end{array}\right.\]

Then, we can construct two daemons $d_1$ and $d_2$ in the following way:
\[\left\{\begin{array}{c}
\forall\pi\in\Pi\setminus\{\pi_1\},d_1(\pi)=d(\pi)\\
d_1(\pi_1)=d(\pi_1)\cup\{\sigma_1\}
\end{array}\right.
\mbox{, and }
\left\{\begin{array}{c}
\forall\pi\in\Pi\setminus\{\pi_2\},d_2(\pi)=d(\pi)\\
d_2(\pi_2)=d(\pi_2)\cup\{\sigma_2\}
\end{array}\right.\]

Then, we can deduce that $d_2(\pi_1)\subsetneq d_1(\pi_1)$ and $d_1(\pi_2)\subsetneq d_2(\pi_2)$, which proves that $d_1$ and $d_2$ are not comparable using the binary relation $\preccurlyeq$.
\end{proof}

Another natural intuition is that if $d$ is more powerful than $d'$ and $d$ belong to a particular daemon class, then $d'$ also belongs to this class. This is formally demonstrated in the following.

\begin{proposition}\label{prop:partialorder}
For a given graph $g$, for any daemons $d$ and $d'$ and for any class of daemons $\mathcal{D}(C,B,E,F)$, the following statements hold:
\[\left.\begin{array}{c}
d'\preccurlyeq d\\
d\in \mathcal{D}(C,B,E,F)
\end{array}\right\}\Rightarrow d'\in\mathcal{D}(C,B,E,F)\]
\end{proposition}

\begin{proof}
Let $g$ be a graph, $d$ and $d'$ be two daemons and $\mathcal{D}(C,B,E,F)$ be a class of daemons such that: $d'\preccurlyeq d$ and $d\in \mathcal{D}(C,B,E,F)$.

Assume that $C=k\mbox{-}\mathcal{C}$ with $k\in\{0,\ldots,diam(g)\}$. As $d\in \mathcal{D}(C,B,E,F)=C\cap B\cap E\cap F$, $d\in k\mbox{-}\mathcal{C}$. By definition:
\[\begin{array}{r@{}l}
\forall \pi\in\Pi,\forall \sigma=(\gamma_0,\gamma_1)&(\gamma_1,\gamma_2)\ldots\in d(\pi),\forall i\in\mathbb{N},\forall (u,v)\in V^2,\\
&[u\neq v \wedge u\in Act(\gamma_i,\gamma_{i+1}) \wedge v\in Act(\gamma_i,\gamma_{i+1})]\Rightarrow dist(g,u,v)>k
\end{array}\]

As $d'\preccurlyeq d$, $\forall \pi\in\Pi,d'(\pi)\subseteq d(\pi)$. Then, we obtain:
\[\begin{array}{r@{}l}
\forall \pi\in\Pi,\forall \sigma=(\gamma_0,\gamma_1)&(\gamma_1,\gamma_2)\ldots\in d'(\pi)\subseteq d(\pi),\forall i\in\mathbb{N},\forall (u,v)\in V^2,\\
&[u\neq v \wedge u\in Act(\gamma_i,\gamma_{i+1}) \wedge v\in Act(\gamma_i,\gamma_{i+1})]\Rightarrow dist(g,u,v)>k
\end{array}\]

This implies that $d'\in k\mbox{-}\mathcal{C}=C$. We can prove in a similar way that $d'\in B$, $d'\in E$ and $d'\in F$. Consequently, we obtain that $d'\in C\cap B\cap E \cap F=\mathcal{D}(C,B,E,F)$, which proves the result.
\end{proof}

\subsection{Preserving execution properties}

Meaningful distributed protocols provide non-trivial properties when operated. A property can be defined as a predicate on computations, valued with \emph{true} when the predicate is satisfied and \emph{false} otherwise. A distributed protocol satisfies a property if its every executions satisfy the corresponding predicate. Conversely, a property is impossible to satisfy if no protocol is such that any of its executions satisfies the corresponding predicate. Formal definitions follow.

\begin{definition}[Execution property]
For a given graph $g$, a property of execution $p$ is a function that associates to each execution a Boolean value.
\[\begin{array}{ccrcl}
p & : & \Sigma_\Pi & \longrightarrow & \{true,false\}\\
  &   & \sigma     & \longmapsto     & p(\sigma)\in\{true,false\}
\end{array}\]
\end{definition}

\begin{definition}[Property satisfaction]
For a given graph $g$, a distributed protocol $\pi$ satisfies a property of execution $p$ under a daemon $d$ (denoted by $\pi\overset{d}{\models} p$) if and only if $\forall \sigma\in d(\pi),p(\sigma)=true$.
\end{definition}

\begin{definition}[Property impossibility]
For a given graph $g$, it is impossible to satisfy a property of execution $p$ under a daemon $d$ (denoted by $d\not\models p$) if and only if $\forall \pi\in\Pi,\exists \sigma\in d(\pi), p(\sigma)=false$.
\end{definition}

The ``more powerful'' meaning that is associated to the $\preccurlyeq$ relation permits to intuitively understand the two following theorems. If a property is guaranteed by a protocol under a daemon $d$, it is also guaranteed using the same protocol under any ``less powerful'' daemon $d'$ (the executions allowed by $d'$ are a -- possibly strict -- subset of those allowed by $d$). Similarly, if a property cannot be guaranteed by any protocol under a daemon $d$, it is also impossible to guarantee this property under a ``more powerful'' daemon $d'$ (the executions that falsifies the property in these allowed by $d$ are also present in those allowed by $d'$). A formal treatment follows.

\begin{theorem}
For a given graph $g$, let $p$ be a property of execution satisfied by a distributed protocol $\pi$ under a daemon $d$. Then,
\[\forall d'\in\mathcal{D}, d'\preccurlyeq d\Rightarrow \pi\overset{d'}{\models} p\]
\end{theorem}

\begin{proof}
Let $g$ be a graph, $p$ be a property of execution satisfied by a distributed protocol $\pi_1$ under a daemon $d$. By definition:
\[\forall \sigma\in d(\pi_1),p(\sigma)=true\]

Assume now that $d'$ is a daemon such that $d'\preccurlyeq d$. By definition:
\[\forall \pi\in\Pi, d'(\pi)\subseteq d(\pi)\]

Consequently, we get:
\[\forall \sigma\in \Sigma_{\pi_1}, \sigma\in d'(\pi_1)\Rightarrow \sigma\in d(\pi_1)\Rightarrow p(\sigma)=true\]

By definition, we obtain that: $\pi_1\overset{d'}{\models} p$, which proves the theorem.
\end{proof}

\begin{theorem}
For a given graph $g$, let $p$ be a property of execution impossible under a daemon $d$. Then,
\[\forall d'\in\mathcal{D}, d\preccurlyeq d'\Rightarrow d'\not\models p\]
\end{theorem}

\begin{proof}
Let $g$ be a graph, $p$ be a property of execution impossible under a daemon $d$. By definition:
\[\forall \pi\in\Pi,\exists \sigma\in d(\pi),p(\sigma)=false\]

Assume now that $d'$ is a daemon such that $d\preccurlyeq d'$. By definition:
\[\forall \pi\in\Pi, d(\pi)\subseteq d'(\pi)\]

Consequently, we obtain:
\[\forall \pi\in\Pi,\exists \sigma\in d(\pi)\subseteq d'(\pi),p(\sigma)=false\]

By definition, we obtain that: $d'\not\models p$, which proves the theorem.
\end{proof}

A less obvious result shows that dealing with canonical daemons (rather than with the classes they represent) is sufficient for comparison purposes. The two derived corollaries demonstrate that using characteristic daemons is also valid for proving properties (or lack hereof) executions. This is formalized in the sequel.

\begin{theorem}\label{th:canonical}
For a given graph $g$, let $d(C,B,E,F)$ and $d(C',B',E',F')$ be two canonical daemons. Then,
\[d(C,B,E,F)\preccurlyeq d(C',B',E',F')\Leftrightarrow 
\left\{\begin{array}{l}
C\subseteq C'\\
B\subseteq B'\\
E\subseteq E'\\
F\subseteq F'\\
\end{array}\right.\]
\end{theorem}

\begin{proof}
We first prove the ``$\Leftarrow$'' part of the theorem.

Assume that there exist a graph $g$ and two canonical daemons $d(C,B,E,F)$ and $d(C',B',E',F')$ such that:
\[\left\{\begin{array}{l}
C\subseteq C'\\
B\subseteq B'\\
E\subseteq E'\\
F\subseteq F'\\
\end{array}\right.\]

We can deduce that $C\cap B\cap E\cap F\subseteq C'\cap B'\cap E'\cap F'$. Then, by the definition of a class of daemons:
\[\mathcal{D}(C,B,E,F)\subseteq \mathcal{D}(C',B',E',F')\]

By the definition of a canonical daemon, $d(C,B,E,F)\in\mathcal{D}(C,B,E,F)$. Hence:
\[d(C,B,E,F)\in\mathcal{D}(C',B',E',F')\]

As $d(C',B',E',F')$ is the canonical daemon of the class $\mathcal{D}(C',B',E',F')$, by definition:
\[\forall \pi\in\Pi,\forall \sigma\in\Sigma_\pi, \sigma\in d(C,B,E,F)(\pi)\Rightarrow \sigma\in d(C',B',E',F')(\pi)\]

In other words,
\[\forall \pi\in\Pi, d(C,B,E,F)(\pi)\subseteq d(C',B',E',F')(\pi)\]

This means that: $d(C,B,E,F)\preccurlyeq d(C',B',E',F')$, which ends the first part of the proof.

Then, we prove the ``$\Rightarrow$'' part of the theorem.

Assume that there exist a graph $g$ and two canonical daemons $d(C,B,E,F)$ and $d(C',B',E',F')$ such that: $d(C,B,E,F)\preccurlyeq d(C',B',E',F')$.

By definition of the $\preccurlyeq$ relation,
\[\forall\pi\in\Pi, d(C,B,E,F)(\pi)\subseteq d(C',B',E',F')(\pi)\]

Let $d$ be a daemon of $\mathcal{D}(C,B,E,F)$. As $d(C,B,E,F)$ is the canonical daemon of the class of daemons $\mathcal{D}(C,B,E,F)$,
\[\begin{array}{rcl}
\forall\pi\in\Pi, \forall \sigma\in\Sigma_\pi, \sigma\in d(\pi) & \Rightarrow & \sigma\in d(C,B,E,F)(\pi)\\
& \Rightarrow & \sigma\in d(C',B',E',F')(\pi)\end{array}\]

In other words, $\forall\pi\in\Pi, d(\pi)\subseteq d(C',B',E',F')(\pi)$. By the definition of the $\preccurlyeq$ relation, this implies that:
\[\forall d\in \mathcal{D}(C,B,E,F), d\preccurlyeq d(C',B',E',F')\]

As $d(C',B',E',F')$ is the canonical daemon of the class of daemons $\mathcal{D}(C',B',E',F')$, $d(C',B',E',F')\in \mathcal{D}(C',B',E',F')$ and Proposition \ref{prop:partialorder} implies
\[\forall d\in\mathcal{D}(C,B,E,F), d\in \mathcal{D}(C',B',E',F')\]

In other words, $C\cap B\cap E\cap F=\mathcal{D}(C,B,E,F)\subseteq \mathcal{D}(C',B',E',F')=C'\cap B'\cap E'\cap F'$. 

Assume by contradiction that $C'\subsetneq C$. By the properties of boundedness, enabledness and fairness (see propositions of Section 3), $(C\setminus C')\cap B\cap E\cap F\neq \emptyset$. So, there exists a daemon $d$ such that $d\in C\cap B\cap E\cap F$ and $d\notin C'$. Then, we can deduce that $d\notin C'\cap B'\cap E'\cap F'$, which contradicts  $C\cap B\cap E\cap F\subseteq C'\cap B'\cap E'\cap F'$. 

By the same way, we can prove that:
\[\left\{\begin{array}{l}
C\subseteq C'\\
B\subseteq B'\\
E\subseteq E'\\
F\subseteq F'\\
\end{array}\right.\]

This result ends the proof.
\end{proof}

\begin{corollary}
For a given graph $g$, let $d(C,B,E,F)$ and $d(C',B',E',F')$ be two canonical daemons. Then, for any property of execution $p$ satisfied by a distributed protocol $\pi$ under $d(C,B,E,F)$, the following statements hold:
\[\left.\begin{array}{l}
C'\subseteq C\\
B'\subseteq B\\
E'\subseteq E\\
F'\subseteq F\\
\end{array}\right\}\Rightarrow \pi\overset{d(C',B',E',F')}{\models} p\]
\end{corollary}

\begin{proof}
This result is a direct corollary from Theorems 1 and 3.
\end{proof}

\begin{corollary}
For a given graph $S$, let $d(C,B,E,F)$ and $d(C',B',E',F')$ be two canonical daemons. Then, for any property of execution $p$ impossible under $d(C,B,E,F)$, the following statements hold:
\[\left.\begin{array}{l}
C\subseteq C'\\
B\subseteq B'\\
E\subseteq E'\\
F\subseteq F'\\
\end{array}\right\}\Rightarrow d(C',B',E',F')\not\models p\]
\end{corollary}

\begin{proof}
This result is a direct corollary from Theorems 2 and 3.
\end{proof}

\subsection{The Case of the Synchronous Daemon}

Although we did not describe it in the previous sections, the \emph{synchronous} daemon play a very important part in the self-stabilization literature. First introduced by Herman~\cite{H90j} to enable analytical tractability of probabilistic self-stabilizing protocols, it was later used in a number of works, either to demonstrate impossibility results (due to initial symmetry~\cite{GT00c}) or to enable efficient solution to existing problems (due to the single scheduling generated~\cite{DHT04ca}). A synchronous daemon simply executes every enabled process at every step. A formal definition follows.

\begin{definition}[Synchronous Daemon]
Given a graph $g$, the synchronous daemon (denoted by $sd$) is defined by:
\[
\forall\pi\in\Pi, \forall \sigma=(\gamma_0,\gamma_1)(\gamma_1,\gamma_2)\ldots \in sd(\pi), \forall i\in\mathbb{N},\forall v\in V, v\in Ena(\gamma_i,\pi)\Rightarrow v\in Act((\gamma_i,\gamma_{i+1}))
\]
\end{definition}

We first show that there is a connection between enabledness and synchrony. Indeed a synchronous daemon cannot prevent an enabled process from being activated, even for a single step.

\begin{proposition}\label{prop:0E=SD}
For any given graph $g$, $0\mbox{-}\mathcal{E}=\{sd\}$.
\end{proposition}

\begin{proof}
Let $g$ be a graph and $d$ be a daemon such that $d\in 0\mbox{-}\mathcal{E}$. We now prove that $d=sd$.

By definition:
\[\begin{array}{r@{}l}
\exists k\in\mathbb{N},\forall \pi\in\Pi,&\forall \sigma=(\gamma_0,\gamma_1)(\gamma_1,\gamma_2)\ldots\in d(\pi),\forall (i,j)\in\mathbb{N}^2,\forall v\in V,\\
& \big[
[v\in Act(\gamma_i,\gamma_{i+1})\wedge(\forall l\in\mathbb{N},l<i\Rightarrow v\notin Act(\gamma_l,\gamma_{l+1}))]\\
& \Rightarrow |\{l\in\mathbb{N}|l<i\wedge v\in Ena(\gamma_l,\pi)\}|=0
\big]\wedge\\
& \big[
[i<j \wedge v\in Act(\gamma_i,\gamma_{i+1}) \wedge v\in Act(\gamma_j,\gamma_{j+1}) \wedge (\forall l\in\mathbb{N},i<l<j\Rightarrow v\notin Act(\gamma_l,\gamma_{l+1}))]\\
& \Rightarrow |\{l\in\mathbb{N}|i<l<j\wedge v\in Ena(\gamma_l,\pi)\}|=0
\big]\wedge\\
& \big[
[v\in Act(\gamma_i,\gamma_{i+1})\wedge(\forall l\in\mathbb{N},l>i\Rightarrow v\notin Act(\gamma_l,\gamma_{l+1}))]\\
& \Rightarrow |\{l\in\mathbb{N}|l>i\wedge v\in Ena(\gamma_l,\pi)\}|=0\big]
\end{array}\]

In other words, no action $(\gamma,\gamma')$ of any execution of $d(\pi)$ for any distributed protocol $\pi$ can satisfy: $\exists v\in V,v\in Ena(\gamma,\pi)\wedge v\notin Act(\gamma,\gamma')$. Hence:
\[\forall\pi\in\Pi,\forall \sigma=(\gamma_0,\gamma_1)(\gamma_1,\gamma_2)\ldots\in d(\pi),\forall i\in\mathbb{N}, \forall v\in V, v\notin Ena(\gamma_i,\pi)\vee v\in Act(\gamma_i,\gamma_{i+1})\]

As $v\in Act(\gamma_i,\gamma_{i+1})$ implies that $v\in Ena(\gamma_i,\pi)$, this property is equivalent to the following:
\[\forall\pi\in\Pi,\forall \sigma=(\gamma_0,\gamma_1)(\gamma_1,\gamma_2)\ldots\in d(\pi),\forall i\in\mathbb{N},\forall v\in V, v\in Ena(\gamma_i,\pi)\Rightarrow v\in Act(\gamma_i,\gamma_{i+1})\]

By the definition of the synchronous daemon, this means that $d=sd$, which ends the proof.
\end{proof}

It may first come to a surprise that boundedness is absolutely not related to synchrony, but as we pointed out previously, boundedness is also not related to fairness. The exact characteristics of the synchronous daemon are captured by the following proposition.

\begin{proposition}
Given a graph $g$, $\mathcal{D}(0\mbox{-}\mathcal{C},\mathcal{D},0\mbox{-}\mathcal{E},\mathcal{SF})$ is the minimal class of $SD$. Moreover, $SD=d(0\mbox{-}\mathcal{C},\mathcal{D},0\mbox{-}\mathcal{E},\mathcal{SF})$.
\end{proposition}

\begin{proof}
First, we prove that $sd\in 0\mbox{-}\mathcal{C}\setminus 1\mbox{-}\mathcal{C}$. By definition, $sd\in 0\mbox{-}\mathcal{C}=\mathcal{D}$. By contradiction, assume that $sd\in 1\mbox{-}\mathcal{C}$. Let $\pi\in\Pi$ be a distributed protocol such that:
\[\exists (\gamma,\gamma')\in\pi, Ena(\gamma,\pi)=V\]

Then, by definition of the synchronous daemon, the first action of any execution $\sigma=(\gamma_0,\gamma_1)(\gamma_1,\gamma_2)\ldots\in sd(\pi)$ starting from $\gamma_0=\gamma$ satisfies: $Act(\gamma_0,\gamma_1)=V$. Consequently, $\sigma$ does not satisfy the property of executions allowed by a $1$-central daemon, which contradicts $sd\in 1\mbox{-}\mathcal{C}$ and proves the result.

Second, we prove that $sd\in\bar{\mathcal{B}}$. As $sd\in\mathcal{D}$, assume for the purpose of contradiction that there exists $k\in\mathbb{N}^*$ such that $sd\in k\mbox{-}\mathcal{B}$. Then, consider a distributed protocol $\pi$ such that:
\[\exists (v,u)\in V^2,\exists (\gamma_0,\ldots,\gamma_{k+3})\in\Gamma^{k+4},\left\{\begin{array}{l}
(\gamma_0,\gamma_1)\in \pi\wedge Ena(\gamma_0,\pi)=\{v\}\\
\forall i\in\{1,\ldots,k+1\},(\gamma_i,\gamma_{i+1})\in \pi\wedge Ena(\gamma_i,\pi)=\{u\}\\
(\gamma_{k+2},\gamma_{k+3})\in \pi\wedge Ena(\gamma_{k+2},\pi)=\{v\}\\
Ena(\gamma_{k+3},\pi)=\emptyset
\end{array}\right.\]

We can observe that the execution $\sigma$ defined by $\sigma=(\gamma_0,\gamma_1)(\gamma_1,\gamma_2)\ldots(\gamma_{k+2},\gamma_{k+3})$ satisfies $\sigma\in sd(\pi)$. But, on the other hand, the following holds:
\[\begin{array}{r@{}l}
\exists\pi\in\Pi,&\exists \sigma=(\gamma_0,\gamma_1)(\gamma_1,\gamma_2)\ldots\in d(\pi),\exists (i=0,j=k+2)\in\mathbb{N}^2,\exists v\in V,\\
& [i<j \wedge v\in Act(\gamma_i,\gamma_{i+1}) \wedge v\in Act(\gamma_j,\gamma_{j+1}) \wedge (\forall l\in\mathbb{N},i<l<j\Rightarrow v\notin Act(\gamma_l,\gamma_{l+1}))]\\
& \wedge \exists u\in V\setminus\{v\},|\{l\in\mathbb{N}|i\leq l<j\wedge u\in Act(\gamma_l,\gamma_{l+1})\}|=k+1
\end{array}\]
 
By the definition of a $k$-bounded daemon, this implies that $sd\notin k\mbox{-}\mathcal{B}$.

We now prove that $sd\in 0\mbox{-}\mathcal{E}$. By Proposition 8, $0\mbox{-}\mathcal{E}=\{sd\}$. This implies that $sd\in 0\mbox{-}\mathcal{E}$.

Next, we prove that $sd\in \mathcal{SF}\setminus\mathcal{GF}$. We start by proving that $sd\in \mathcal{SF}$. By the definition of the synchronous daemon:
\[\forall\pi\in\Pi,\forall \sigma=(\gamma_0,\gamma_1)(\gamma_1,\gamma_2)\ldots\in\Sigma_\pi, \forall v\in V, (\exists i\in\mathbb{N},v\in Ena(\gamma_i,\pi))\Rightarrow v \in Act(\gamma_j,\gamma_{j+1})\]

Consequently,
\[\begin{array}{r@{}l}
\forall\pi\in\Pi,&\forall \sigma=(\gamma_0,\gamma_1)(\gamma_1,\gamma_2)\ldots\in\Sigma_\pi,\\
& [\exists i\in\mathbb{N},\exists v\in V,(\forall j\geq i,\exists k\geq j,v\in Ena(\gamma_k,\pi))\wedge(\forall j\geq i,v\notin Act(\gamma_j,\gamma_{j+1}))]\Rightarrow \sigma\notin sd(\pi)
\end{array}\]

By the definition of a strongly fair daemon, this implies that $sd\in \mathcal{SF}$. Now, we prove that $sd\notin\mathcal{GF}$. Consider a distributed protocol $\pi$ such that:
\[\exists (\gamma,\gamma',\gamma'')\in \Gamma^3,
\left\{\begin{array}{l}
(\gamma,\gamma')\in \pi\wedge Act(\gamma,\gamma')\subsetneq Ena(\gamma,\pi)\\
(\gamma,\gamma'')\in \pi\wedge Act(\gamma,\gamma'')=Ena(\gamma,\pi)\\
(\gamma'',\gamma)\in \pi\wedge Act(\gamma'',\gamma)=Ena(\gamma'',\pi)
\end{array}\right.\]

We can construct an execution $\sigma$ of $\pi$ starting from $\gamma$ in the following way: $\sigma=(\gamma,\gamma'')(\gamma'',\gamma)(\gamma,\gamma'')\ldots$. We can observe that $\sigma\in sd(\pi)$ (since at each action, any enabled vertex is activated). Consequently,
\[\begin{array}{r@{}l}
\exists\pi\in\Pi,&\exists \sigma=(\gamma_0,\gamma_1)(\gamma_1,\gamma_2)\ldots\in sd(\pi),\exists (\gamma,\gamma')\in \pi,\\
& \exists i=0\in\mathbb{N},(\forall j\geq i,\exists k=2j\geq j,\gamma_k=\gamma)\wedge(\forall j\geq i,(\gamma_j,\gamma_{j+1})\neq (\gamma,\gamma'))
\end{array}\]

By the definition of a Gouda fair daemon, this implies that $sd\notin\mathcal{GF}$.

The four previous results imply that $\mathcal{D}(0\mbox{-}\mathcal{C},\mathcal{D},0\mbox{-}\mathcal{E},\mathcal{SF})$ is the minimal class of $sd$. As $\mathcal{D}(0\mbox{-}\mathcal{C},\mathcal{D},0\mbox{-}\mathcal{E},\mathcal{SF})\subseteq 0\mbox{-}\mathcal{E}$ by definition and $0\mbox{-}\mathcal{E}=\{sd\}$ by Proposition \ref{prop:0E=SD}, we can deduce that $\mathcal{D}(0\mbox{-}\mathcal{C},\mathcal{D},0\mbox{-}\mathcal{E},\mathcal{SF})=\{sd\}$. Then, the definition of a canonical daemon implies that $sd=d(0\mbox{-}\mathcal{C},\mathcal{D},0\mbox{-}\mathcal{E},\mathcal{SF})$, which completes the proof.
\end{proof}

\subsection{A map of classical daemons}

We are now ready to present our map for ``classical'' daemons (\emph{i.e.} daemons most frequently used in the literature. Using our taxonomy, these daemons can be defined as follows.

\begin{definition}[Classical daemons]
Given a graph $g$, the classical daemons of the literature are defined as follows:
\begin{itemize}
\item The unfair daemon (denoted by $ufd$) is $d(\mathcal{D},\mathcal{D},\mathcal{D},\mathcal{D})$.
\item The weakly fair daemon (denoted by $wfd$) is $d(\mathcal{D},\mathcal{D},\mathcal{D},\mathcal{WF})$.
\item The strongly fair daemon (denoted by $sfd$) is $d(\mathcal{D},\mathcal{D},\mathcal{D},\mathcal{SF})$.
\item The Gouda fair daemon (denoted by $gfd$) is $d(\mathcal{D},\mathcal{D},\mathcal{D},\mathcal{GF})$.
\item The locally central unfair daemon (denoted by $1\mbox{-}ufd$) is $d(1\mbox{-}\mathcal{C},\mathcal{D},\mathcal{D},\mathcal{D})$.
\item The locally central weakly fair daemon (denoted by $1\mbox{-}wfd$) is $d(1\mbox{-}\mathcal{C},\mathcal{D},\mathcal{D},\mathcal{WF})$.
\item The locally central strongly fair daemon (denoted by $1\mbox{-}sfd$) is $d(1\mbox{-}\mathcal{C},\mathcal{D},\mathcal{D},\mathcal{SF})$.
\item The locally central Gouda fair daemon (denoted by $1\mbox{-}gfd$) is $d(1\mbox{-}\mathcal{C},\mathcal{D},\mathcal{D},\mathcal{GF})$.
\item The central unfair daemon (denoted by $0\mbox{-}ufd$) is $d(0\mbox{-}\mathcal{C},\mathcal{D},\mathcal{D},\mathcal{D})$.
\item The central weakly fair daemon (denoted by $0\mbox{-}wfd$) is $d(0\mbox{-}\mathcal{C},\mathcal{D},\mathcal{D},\mathcal{WF})$.
\item The central strongly fair daemon (denoted by $0\mbox{-}sfd$) is $d(0\mbox{-}\mathcal{C},\mathcal{D},\mathcal{D},\mathcal{SF})$.
\item The central Gouda fair daemon (denoted by $0\mbox{-}gfd$) is $d(0\mbox{-}\mathcal{C},\mathcal{D},\mathcal{D},\mathcal{GF})$.
\end{itemize}
\end{definition}

\begin{figure}[htbp]
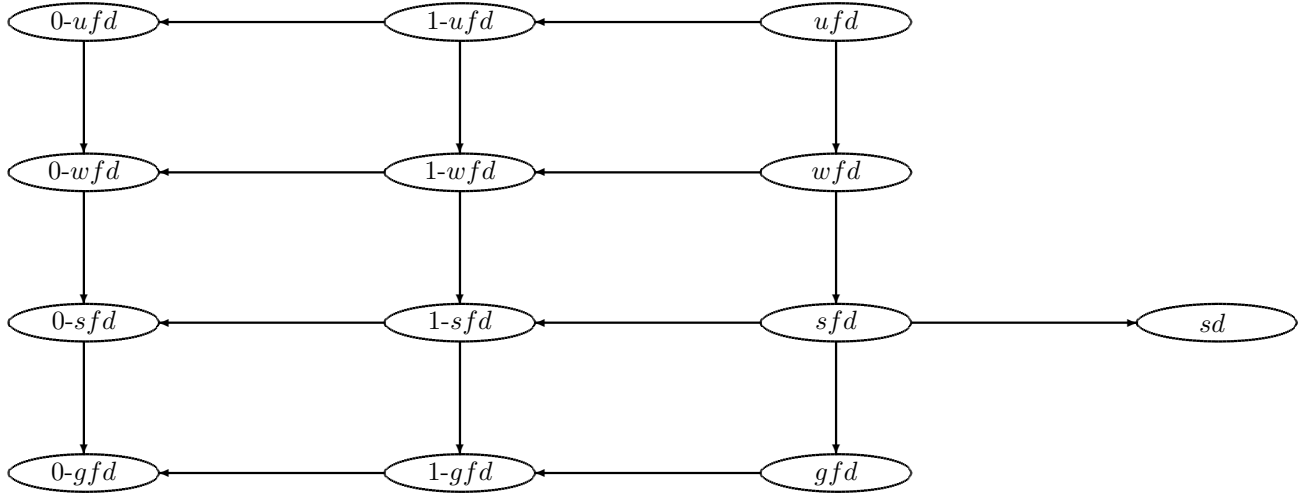

\noindent \begin{centering} \include{lfig/ClassicalDaemons}
  \par\end{centering}
 \caption{Relationship between classical daemons (an arrow from a daemon $d$ to a daemon $d'$ means that $d'\preccurlyeq d$, note that we remove all arrows obtained by transitivity).}
\label{fig:classicalDaemons}
\end{figure}

Now, our main theorem (Theorem~\ref{th:canonical}) permits to map the relationships between all classical daemons in the literature is a rather compact format. For any given graph $g$, Figure~\ref{fig:classicalDaemons} depicts graphically those relationships.

\section{Daemon Transformers}\label{sec:transformers}

As it is easier to write distributed protocols under daemons providing strong properties (that is, under weak daemons that allow only a limited set of possible executions, such as a central or a bounded daemon), many authors provide protocols to simulate the operation of a weak daemon under a strong one. Such protocols are called \emph{daemon transformers}. Note that several works in the area of self-stabilization may be used as daemon transformers although they were not initially designed with this goal in mind (\emph{e.g.} a self-stabilizing token circulation protocol that performs under the unfair distributed daemon can easily be turned into a daemon transformer that provides a central daemon out of an unfair distributed one).

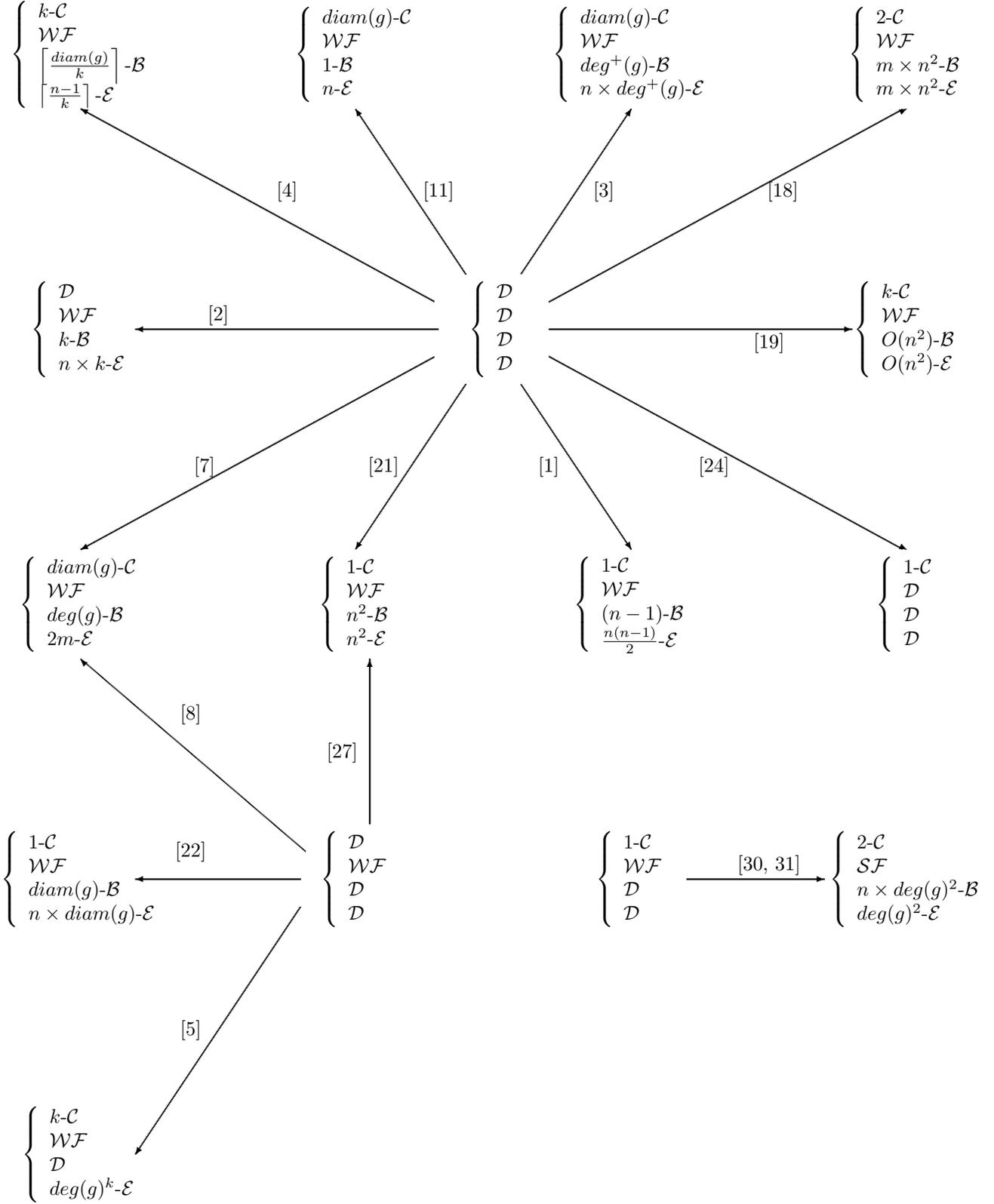
\begin{figure}[htb]
\noindent \begin{centering} \ifx\JPicScale\undefined\def\JPicScale{0.97}\fi
\unitlength \JPicScale mm
\begin{picture}(152.5,202.5)(0,0)
\put(77.5,152.5){\makebox(0,0)[cc]{
$\left\{\begin{array}{l}
\mathcal{D}\\
\mathcal{D}\\
\mathcal{D}\\
\mathcal{D}
\end{array}\right.$}}

\put(2.5,202.5){\makebox(0,0)[cc]{
$\left\{\begin{array}{l}
k\mbox{-}\mathcal{C}\\
\mathcal{WF}\\
\left\lceil\frac{diam(g)}{k}\right\rceil\mbox{-}\mathcal{B}\\
\left\lceil\frac{n-1}{k}\right\rceil\mbox{-}\mathcal{E}
\end{array}\right.$}}

\put(40,177.5){\makebox(0,0)[cc]{\cite{BP08c}}}

\linethickness{0.3mm}
\put(12.5,52.5){\line(1,0){30}}
\put(12.5,52.5){\vector(-1,0){0.12}}
\linethickness{0.3mm}
\multiput(2.5,92.5)(0.14,-0.12){292}{\line(1,0){0.14}}
\put(2.5,92.5){\vector(-1,1){0.12}}
\linethickness{0.3mm}
\multiput(12.5,2.5)(0.12,0.18){250}{\line(0,1){0.18}}
\put(12.5,2.5){\vector(-2,-3){0.12}}
\linethickness{0.3mm}
\put(112.5,52.5){\line(1,0){25}}
\put(137.5,52.5){\vector(1,0){0.12}}
\linethickness{0.3mm}
\put(87.5,152.5){\line(1,0){55}}
\put(142.5,152.5){\vector(1,0){0.12}}
\linethickness{0.3mm}
\multiput(82.5,162.5)(0.12,0.18){167}{\line(0,1){0.18}}
\put(102.5,192.5){\vector(2,3){0.12}}
\linethickness{0.3mm}
\multiput(52.5,192.5)(0.12,-0.18){167}{\line(0,-1){0.18}}
\put(52.5,192.5){\vector(-2,3){0.12}}
\linethickness{0.3mm}
\multiput(52.5,112.5)(0.12,0.18){167}{\line(0,1){0.18}}
\put(52.5,112.5){\vector(-2,-3){0.12}}
\linethickness{0.3mm}
\multiput(82.5,142.5)(0.12,-0.18){167}{\line(0,-1){0.18}}
\put(102.5,112.5){\vector(2,-3){0.12}}
\linethickness{0.3mm}
\multiput(87.5,147.5)(0.22,-0.12){292}{\line(1,0){0.22}}
\put(152.5,112.5){\vector(2,-1){0.12}}
\linethickness{0.3mm}
\multiput(87.5,157.5)(0.22,0.12){292}{\line(1,0){0.22}}
\put(152.5,192.5){\vector(2,1){0.12}}
\linethickness{0.3mm}
\multiput(2.5,192.5)(0.22,-0.12){292}{\line(1,0){0.22}}
\put(2.5,192.5){\vector(-2,1){0.12}}
\linethickness{0.3mm}
\multiput(2.5,112.5)(0.22,0.12){292}{\line(1,0){0.22}}
\put(2.5,112.5){\vector(-2,-1){0.12}}
\linethickness{0.3mm}
\put(12.5,152.5){\line(1,0){55}}
\put(12.5,152.5){\vector(-1,0){0.12}}

\put(52.5,52.5){\makebox(0,0)[cc]{
$\left\{\begin{array}{l}
\mathcal{D}\\
\mathcal{WF}\\
\mathcal{D}\\
\mathcal{D}
\end{array}\right.$}}

\put(2.5,52.5){\makebox(0,0)[cc]{
$\left\{\begin{array}{l}
1\mbox{-}\mathcal{C}\\
\mathcal{WF}\\
diam(g)\mbox{-}\mathcal{B}\\
n\times diam(g)\mbox{-}\mathcal{E}
\end{array}\right.$}}

\put(22.5,57.5){\makebox(0,0)[cc]{\cite{GH07j}}}

\put(52.5,202.5){\makebox(0,0)[cc]{
$\left\{\begin{array}{l}
diam(g)\mbox{-}\mathcal{C}\\
\mathcal{WF}\\
1\mbox{-}\mathcal{B}\\
n\mbox{-}\mathcal{E}
\end{array}\right.$}}

\put(67.5,177.5){\makebox(0,0)[cc]{\cite{D74j}}}

\put(2.5,102.5){\makebox(0,0)[cc]{
$\left\{\begin{array}{l}
diam(g)\mbox{-}\mathcal{C}\\
\mathcal{WF}\\
deg(g)\mbox{-}\mathcal{B}\\
2m\mbox{-}\mathcal{E}
\end{array}\right.$}}

\put(22.5,82.5){\makebox(0,0)[cc]{\cite{DJPV00j}}}

\put(25,127.5){\makebox(0,0)[cc]{\cite{DGT04j}}}

\put(102.5,202.5){\makebox(0,0)[cc]{
$\left\{\begin{array}{l}
diam(g)\mbox{-}\mathcal{C}\\
\mathcal{WF}\\
deg^+(g)\mbox{-}\mathcal{B}\\
n\times deg^+(g)\mbox{-}\mathcal{E}
\end{array}\right.$}}

\put(97.5,177.5){\makebox(0,0)[cc]{\cite{BGJD02j}}}

\put(2.5,2.5){\makebox(0,0)[cc]{
$\left\{\begin{array}{l}
k\mbox{-}\mathcal{C}\\
\mathcal{WF}\\
\mathcal{D}\\
deg(g)^k\mbox{-}\mathcal{E}
\end{array}\right.$}}

\put(22.5,25){\makebox(0,0)[cc]{\cite{DNT09j}}}

\put(152.5,202.5){\makebox(0,0)[cc]{
$\left\{\begin{array}{l}
2\mbox{-}\mathcal{C}\\
\mathcal{WF}\\
m\times n^2\mbox{-}\mathcal{B}\\
m\times n^2\mbox{-}\mathcal{E}
\end{array}\right.$}}

\put(130,177.5){\makebox(0,0)[cc]{\cite{GGHKM04j}}}

\put(152.5,152.5){\makebox(0,0)[cc]{
$\left\{\begin{array}{l}
k\mbox{-}\mathcal{C}\\
\mathcal{WF}\\
O(n^2)\mbox{-}\mathcal{B}\\
O(n^2)\mbox{-}\mathcal{E}
\end{array}\right.$}}

\put(127.5,150){\makebox(0,0)[cc]{\cite{GHJT08j}}}

\put(152.5,102.5){\makebox(0,0)[cc]{
$\left\{\begin{array}{l}
1\mbox{-}\mathcal{C}\\
\mathcal{D}\\
\mathcal{D}\\
\mathcal{D}
\end{array}\right.$}}

\put(117.5,127.5){\makebox(0,0)[cc]{\cite{GT07cb}}}

\put(102.5,102.5){\makebox(0,0)[cc]{
$\left\{\begin{array}{l}
1\mbox{-}\mathcal{C}\\
\mathcal{WF}\\
(n-1)\mbox{-}\mathcal{B}\\
\frac{n(n-1)}{2}\mbox{-}\mathcal{E}
\end{array}\right.$}}

\put(87.5,127.5){\makebox(0,0)[cc]{\cite{BDGM02j}}}

\put(52.5,102.5){\makebox(0,0)[cc]{
$\left\{\begin{array}{l}
1\mbox{-}\mathcal{C}\\
\mathcal{WF}\\
n^2\mbox{-}\mathcal{B}\\
n^2\mbox{-}\mathcal{E}
\end{array}\right.$}}

\put(57.5,127.5){\makebox(0,0)[cc]{\cite{GH97c}}}

\put(2.5,152.5){\makebox(0,0)[cc]{
$\left\{\begin{array}{l}
\mathcal{D}\\
\mathcal{WF}\\
k\mbox{-}\mathcal{B}\\
n\times k\mbox{-}\mathcal{E}
\end{array}\right.$}}

\put(27.5,155){\makebox(0,0)[cc]{\cite{BGJ01c}}}

\put(102.5,52.5){\makebox(0,0)[cc]{
$\left\{\begin{array}{l}
1\mbox{-}\mathcal{C}\\
\mathcal{WF}\\
\mathcal{D}\\
\mathcal{D}
\end{array}\right.$}}

\put(152.5,52.5){\makebox(0,0)[cc]{
$\left\{\begin{array}{l}
2\mbox{-}\mathcal{C}\\
\mathcal{SF}\\
n\times deg(g)^2\mbox{-}\mathcal{B}\\
deg(g)^2\mbox{-}\mathcal{E}
\end{array}\right.$}}

\put(127.5,55){\makebox(0,0)[cc]{\cite{K01j,K05j}}}

\linethickness{0.3mm}
\put(55,62.5){\line(0,0){30}}
\put(55,92.5){\vector(0,0){0.12}}

\put(50,75.5){\makebox(0,0)[cc]{\cite{JADT02j}}}

\end{picture}
  \par\end{centering}
 \caption{Summary of existing daemon transformers.}
\label{fig:transformers}
\end{figure}

In the following, we propose a survey of the main daemon transformers that also preserve the property of self-stabilization. That is, the protocol transforming the daemon is a self-stabilizing one. Figure~\ref{fig:transformers} summarizes this survey and maps for each daemon transformer the initial daemon and the simulated one. We restrict ourselves to deterministic daemon transformers in order to be able to exactly compute the characteristics of the simulated daemon. Note that features of the emulated daemon (centrality, fairness, boundedness, and enabledness) provided in the following are satisfied only after the stabilization of the daemon transformer. In the sequel, we use the notation $d\longmapsto d'$ to denote that a daemon transformer simulates $d'$ while operating under $d$.

\paragraph{Alternator-based daemon transformers.} In 1997, Gouda and Haddix~\cite{GH97c} introduced the alternator problem. Roughly speaking, the aim is to design a protocol such that no neighbors are enabled simultaneously yet ensures that some fairness property holds (namely, between any two steps of a particular process, any of its neighbors may execute at most one step). They claim that this protocol is useful to simulate a locally central daemon under a distributed one. Actually, this protocol ensures the following daemon transformation: $ufd \longmapsto d(1\mbox{-}\mathcal{C},\mathcal{WF},n^2\mbox{-}\mathcal{B},n^2\mbox{-}\mathcal{E})$ and works on chain topologies only. Johnen \emph{et al.} \cite{JADT02j} later designed an alternator for any oriented tree but require the initial daemon to be weakly fair. In other words, they provide the following daemon transformer: $wfd \longmapsto d(1\mbox{-}\mathcal{C},\mathcal{WF},n^2\mbox{-}\mathcal{B},n^2\mbox{-}\mathcal{E})$. Finally, Gouda and Haddix~\cite{GH07j} provided an alternator for an arbitrary underlying communication graph that provides the following daemon transformation: $wfd \longmapsto d(1\mbox{-}\mathcal{C},\mathcal{WF},diam(g)\mbox{-}\mathcal{B},(n\times diam(g))\mbox{-}\mathcal{E})$. This last transformer makes the following assumption: the graph is identified (that is, every vertex has a unique identifier) and each vertex knows the cyclic distance of the graph (the cyclic distance is defined as the number of edges of the longest simple cycle if the graph has cycles, and two otherwise).

\paragraph{Mutual exclusion-based daemon transformers.} The classical mutual exclusion problem requires that no two vertices are simultaneously in critical section and that every vertex infinitely often enters critical section. So, any self-stabilizing mutual exclusion protocol may be turned into a daemon transformer that provides a central weakly fair daemon. In his seminal work on self-stabilization~\cite{D74j}, Dijkstra proposed a self-stabilizing mutual exclusion protocol for ring topologies (using a token circulation) under a distributed unfair daemon. His protocol needs however a distinguished vertex (that is, one vertex executes a protocol that is different from every other). Formally, we can derive the following daemon transformation from this protocol: $ufd\longmapsto d(diam(g)\mbox{-}\mathcal{C},\mathcal{WF},1\mbox{-}\mathcal{B},n\mbox{-}\mathcal{E})$. From this first protocol, several works later revisited the mutual exclusion problem. From a daemon transformation viewpoint, the most interesting ones follow. Using a token circulation, Beauquier \emph{et al.} (\cite{BGJD02j}) provide a $ufd\longmapsto d(diam(g)\mbox{-}\mathcal{C},\mathcal{WF},deg^+(g)\mbox{-}\mathcal{B},n\times deg^+(g)\mbox{-}\mathcal{E})$ daemon transformation on oriented graph whenever the graph is strongly connected. Still on graphs with a distinguished vertex, Datta \emph{et al.} provided~\cite{DJPV00j} a self-stabilizing depth-first token circulation that perform the following daemon transformation: $wfd\longmapsto d(diam(g)\mbox{-}\mathcal{C},\mathcal{WF},deg(g)\mbox{-}\mathcal{B},2m\mbox{-}\mathcal{E})$. Finally, Datta \emph{et al.}~\cite{DGT04j} improved this result enabling the same daemon transformation but starting from an unfair daemon (more formally, they achieve the following daemon transformation: $ufd\longmapsto d(diam(g)\mbox{-}\mathcal{C},\mathcal{WF},deg(g)\mbox{-}\mathcal{B},2m\mbox{-}\mathcal{E})$) and they do not require the existence of a distinguished vertex.

\paragraph{Local mutual exclusion-based daemon transformers.} Local mutual exclusion refines mutual exclusion since it requires the same exclusion and liveness properties but only within a vicinity around each vertex (and not for the whole graph as for the -- global -- mutual exclusion problem). In other words, a $k$-local mutual exclusion protocol ensures that no two vertices are simultaneously in critical section if their distance is less than $k$ and that any vertex enters infinitely often in critical section. Hence, we can easily design a daemon transformer providing a weakly fair $k$-central daemon from such a protocol. Note that the aforementioned alternator protocols solve a particular instance of $1$-local mutual exclusion.

A classical solution to $1$-local mutual exclusion has been proposed by Beauquier \emph{et al.}~\cite{BDGM02j} using unbounded memory at each vertex. This protocol ensures the following daemon transformation: $ufd\longmapsto d(1\mbox{-}\mathcal{C},\mathcal{WF},(n-1)\mbox{-}\mathcal{B},\frac{n(n-1)}{2}\mbox{-}\mathcal{E})$. Using only a bounded memory, Gairing \emph{et al.} provided\cite{GGHKM04j} a $2$-local mutual exclusion that can be turned into a $ufd\longmapsto d(2\mbox{-}\mathcal{C},\mathcal{WF},m\times n^2\mbox{-}\mathcal{B},m\times n^2\mbox{-}\mathcal{E})$ daemon transformer.

Several works give more general solutions dealing with $k$-local mutual exclusion for any integer $k$. For example, Goddart \emph{et al.} generalize \cite{GHJT08j} the work of Gairing \emph{et al.}~\cite{GGHKM04j}. Their solution performs the $ufd\longmapsto d(k\mbox{-}\mathcal{C},\mathcal{WF},O(n^2)\mbox{-}\mathcal{B},O(n^2)\mbox{-}\mathcal{E})$ daemon transformation. Using a local clock synchronization, Boulinier and Petit provide~\cite{BP08c} a wavelets protocol that can be used for $k$-local mutual exclusion. Hence, their protocol gives the following daemon transformation: $ufd\longmapsto d(k\mbox{-}\mathcal{C},\mathcal{WF},\left\lceil\frac{diam(g)}{k}\right\rceil\mbox{-}\mathcal{B},\left\lceil\frac{n-1}{k}\right\rceil\mbox{-}\mathcal{E})$. Danturi \emph{et al.}~\cite{DNT09j} deal with dining philosophers with generic conflicts under a distributed weakly fair daemon. The main idea is to clearly distinguish the communication graph from the conflict graph. If we consider that two vertices are in conflict if they are at distance less than $k$ from each other, this protocol ensures $k$-local mutual exclusion. This protocol provides the $wfd\longmapsto d(k\mbox{-}\mathcal{C},\mathcal{WF},\mathcal{D},deg(g)^k\mbox{-}\mathcal{E})$ daemon transformation but requires each vertex to be the root of a tree spanning its $k$-neighborhood.

Finally, Potop-Butucaru and Tixeuil introduced in \cite{GT07cb} a weaker version of $1$-local mutual exclusion by replacing the fairness property by a progress property. This new problem was called a conflict manager and leads to the $ufd\longmapsto 1\mbox{-}ufd$ daemon transformation. To our knowledge, this daemon transformer is the only one to perform a transformation according to a single identifier daemon characteristic.

Note that all solutions presented in this paragraph require the graph to be identified. 

\paragraph{Other daemon transformers.} Even if they transform several characteristics of daemons (with the notable exception of \cite{GT07cb}), all previously mentioned daemon transformers are designed for transforming only the \emph{distribution} of daemons. Indeed, only a few works dealt with transforming other daemon characteristics.

Regarding fairness transformation, Karaata~\cite{K01j} provided a daemon transformer to perform strong fairness under weak fairness. More formally, this protocol is a $1\mbox{-}wfd\longmapsto d(2\mbox{-}\mathcal{C},\mathcal{SF},n\times deg(g)^2\mbox{-}\mathcal{B},deg(g)^2\mbox{-}\mathcal{E})$ daemon transformer. This protocol needs the graph to be identified and each vertex to have an unbounded memory. Karaata later refined~\cite{K05j} the protocol to perform exactly the same daemon transformation but requiring only an identified graph.

Using cross-over composition, Beauquier \emph{et al.}\cite{BGJ01c} gave a generic transformer for enabledness. More precisely, they design a $ufd\longmapsto d(\mathcal{D},\mathcal{WF},k\mbox{-}\mathcal{B},n\times k\mbox{-}\mathcal{E})$ daemon transformer whenever a transformer that provides $k$-boundedness is available.

\section{Conclusion}\label{sec:conclusion}

We surveyed existing scheduling hypotheses made in the literature in self-stabilization, commonly referred to under the notion of \emph{daemon}. We showed that four main characteristics (distribution, fairness, boundedness, and enabledness) are enough to encapsulate the various differences presented in existing work. Our naming scheme makes it easy to compare daemons of particular classes, and to extend existing possibility or impossibility results to new daemons. We further examined existing daemon transformer schemes and provided the exact transformed characteristics of those transformers in our taxonomy.

Two obvious extensions of this work are to include system hypotheses that are not related to scheduling (\emph{e.g.} atomicity) and to further refine the taxonomy to include recently introduced randomized scheduling~\cite{DTY08c}. 

\bibliographystyle{plain}
\bibliography{biblio}

\end{document}